\documentclass{elsart}
\usepackage{epsfig}
\usepackage[english]{babel}
\usepackage{cite}
\usepackage{multirow}
\usepackage{rotating}

\begin{document}
\begin{frontmatter}

\title{On the universality class of the 3d Ising model with long-range-correlated disorder}
\author[Franko]{D. Ivaneyko\corauthref{cor}},
\corauth[cor]{Corresponding author.}
\ead{ivaneiko@ktf.franko.lviv.ua}
\author[Nancy]{B. Berche},
\ead{berche@lpm.u-nancy.fr}
\author[ICMP,Linz]{Yu. Holovatch},
\ead{hol@icmp.lviv.ua}
\author[ICMP,Potsdam]{J. Ilnytskyi}
\ead{iln@icmp.lviv.ua}

\address[Franko]{Ivan Franko National University of Lviv,
               79005 Lviv, Ukraine}
\address[Nancy]{Laboratoire de Physique des Mat\'eriaux,
               Universit\'e Henri Poincar\'e, Nancy 1,
               54506 Vand\oe uvre les Nancy Cedex, France}
\address[ICMP]{Institute for Condensed Matter Physics,
               National Acad. Sci. of Ukraine,
               79011 Lviv, Ukraine}
\address[Linz]{Institut f\"ur Theoretische Physik,
               Johannes Kepler Universit\"at Linz,
               4040 Linz, Austria}
\address[Potsdam]{Institut f\"ur Physik, Universit\"at Potsdam,
               14469 Potsdam, Deutschland}
\begin{abstract}
We analyze a controversial topic about the universality class of the
three-di\-men\-si\-o\-nal Ising model with long-range-correlated
disorder. Whereas both theoretical and numerical studies agree on the
validity of extended Harris criterion (A.~Weinrib, B.I.~Halperin,
Phys. Rev. B 27 (1983) 413) and indicate the existence of a new
universality class, the numerical values of the critical exponents
found so far differ essentially. To resolve this discrepancy we
perform extensive Monte Carlo simulations of a 3d Ising model with
non-magnetic impurities being arranged in a form of lines along
randomly chosen axes of a lattice. The Swendsen-Wang algorithm is used
alongside with a histogram reweighting technique and the finite-size
scaling analysis to evaluate the values of critical exponents
governing the magnetic phase transition. Our estimates for these
exponents differ from both previous numerical simulations and are in
favour of a non-trivial dependency of the critical exponents on the
peculiarities of long-range correlations decay.
\end{abstract}

\end{frontmatter}

\section{Introduction}\label{I} Critical properties of structurally
disordered magnets remain a problem of great interest in condensed
matter physics, as far as real magnetic crystals are usually
non-ideal. Commonly, in the theoretical studies, as well as in the
MC simulations, one considers point-like uncorrelated quenched
non-magnetic impurities \cite{reviews}. However, in real magnets one
encounters non-idealities of structure, which cannot be modeled by
simple point-like uncorrelated defects. Indeed, magnetic crystals
often contain defects of a more complex structure: linear
dislocations, planar grain boundaries, three-dimensional cavities or
regions of different phases, embedded in the matrix of the original
crystal, as well as various complexes (clusters) of point-like
non-magnetic impurities \cite{defectbook}. Therefore, a challenge is
to offer a consistent description of the critical phenomena
influenced by the presence of such complicated defects.

Different models of structural disorder have arisen as an attempt to
describe such defects. In this paper we concentrate on the so-called
long-range-correlated disorder when the point-like defects are
correlated and the resulting critical behaviour depends on the type of
this correlation. Several models have been proposed for description of
such a dependence \cite{McCoy,Dorogovtsev80,Weinrib83,Lee92}, a
subject of extensive analytical
\cite{Dorogovtsev80,Weinrib83,Lee92,Boyanovsky82,Lawrie84,Yamazaki,%
Fedorenko04,Blavatska,Prudnikov,eta,Vojta} and numerical
\cite{Lee92,Vojta,Ballesteros99,Prudnikov05,Bagamery05} treatment. A
common outcome of the above studies is that although the concentration
of non-magnetic impurities is taken to be far from the percolation
threshold, in the region of weak dilution, the impurities make a
crucial influence on an onset of ordered ferromagnetic phase. Given
that the pure (undiluted) magnet possesses a second-order phase
transition at certain critical temperature $T_c$, an influence of the
weak dilution may range from the decrease of $T_c$ to the changes in
the universality class and even to the smearing off this transition
\cite{Vojta}. Moreover, the critical exponents governing power low
scaling in the vicinity of $T_c$ may depend on the parameters of
impurity-impurity correlation.

To give an example, the Harris criterion, which holds for the
energy-coupled uncorrelated disorder \cite{Harris74} is modified when
the disorder is long-range correlated
\cite{Boyanovsky82,Weinrib83}. In particular, when the
impurity-impurity pair correlation function $g(r)$ decays at large
distances $r$ according to a power law:
\begin{equation} \label{1}
g(r)\sim 1/r^a, \hspace{3em} r \to \infty
\end{equation}
the asymptotic critical exponents governing magnetic phase
transition (and hence the universality class of the transition) do
change if \cite{Weinrib83}
\begin{equation} \label{2}
\nu^{\rm pure} < 2/a,
\end{equation}
where $\nu^{\rm pure}$ is the correlation length critical exponent
of the undiluted magnet. The above condition (\ref{2}) holds for
$a<d$, $d$ being the space (lattice) dimension. For $a>d$ the usual
Harris criterion \cite{Harris74} is recovered and condition
(\ref{2}) is substituted by   $\nu^{\rm pure} < 2/d$.

The fact, that the power of the correlation decay might be a relevant
parameter at $a<d$ can be easily understood observing an asymptotics of
the Fourier transform $g_k$ of $g(r)$ at small wave vector numbers
$k$. From (\ref{1}) one arrives at $g_k\sim k^{a-d}$, which for $a<d$
leads to a singular behaviour at $k\to 0$. As far as the small $k$
region defines the criticality, the systems with $a<d$ are good
candidates to manifest changes in the critical behaviour with respect
to their undiluted counterparts. On contrary, impurity-impurity
correlations at $a>d$ do not produce additional singularities with
respect to the uncorrelated point-like impurities, therefore they are
referred to as the short-range correlated. In turn, the disorder
characterized by Eq.~(\ref{1}) with $a<d$ is called the long-range
correlated.

There are different ways to model systems with the
long-range-correlated disorder governed by Eq. (\ref{1}). The most
direct interpretation relies on the observation that the integer $a$
in Eq. (\ref{1}) corresponds to the large $r$ behaviour of the pair
correlation function for the impurities in the form of points ($a=d$),
lines ($a=d-1$), and planes ($a=d-2$) \cite{Weinrib83}.  Since the
last two objects extend in space, the impurities with $a<d$ sometimes
are called the extended ones. Note that the isotropic form of the pair
correlation function (\ref{1}) demands random orientation of such
spatially extended objects.\footnote{Anisotropic distributions of
extended impurities are treated in Refs.
\cite{McCoy,Dorogovtsev80,Boyanovsky82,Lawrie84,Yamazaki,Blavatska,Vojta}.
} Non-integer $a$ sometimes are treated in terms of a fractal
dimension of impurities, see e.g. \cite{Vasquez}. Besides
energy-coupled disorder, the power-low correlation decay (\ref{1}) is
relevant for the thermodynamic phase transition in random field
systems \cite{Nicolaides00}, percolation \cite{Weinrib84}, scaling of
polymer macromolecules at presence of porous medium
\cite{Blavatska1,Blavatska01}, zero-temperature quantum phase
transitions \cite{Igloi05}.

Our paper was stimulated by the observations of obvious discrepancies
in the state-of-the-art analysis of criticality in three-dimensional
Ising magnets with the long-range-correlated disorder governed by
Eq. (\ref{1}). Indeed, since for the pure $d=3$ Ising model $\nu^{\rm
pure}=0.6304(13)$ \cite{Guida98}, the long-range correlated disorder
should change its universality class according to
Eq.~(\ref{2}). Whereas both theoretical and numerical studies agree on
the validity of extended Harris criterion (\ref{2}) and bring about
the new universality class
\cite{Weinrib83,Prudnikov,Ballesteros99,Prudnikov05}, the numerical
values of the critical exponents being evaluated differ
essentially. We list the values of the exponents found so far by
different approaches in table \ref{tab1} and refer the reader to the
section \ref{II} for a more detailed discussion of this issue. Here,
we would like to point out that presently the results of each of
existing analytical approaches (Refs. \cite{Weinrib83} and
\cite{Prudnikov}) is confirmed by only one numerical simulation
(Refs. \cite{Ballesteros99} and \cite{Prudnikov05}, respectively). To
resolve such a bias, we perform MC simulations of a $d=3$ Ising model
with extended impurities and evaluate critical exponents governing
ferromagnetic phase transition. As it will become evident from the
further account, our estimates for the exponents differ from the
results of two numerical simulations performed so far
\cite{Ballesteros99,Prudnikov05} and are in favour of a non-trivial
dependency of the critical exponents on the peculiarities of
long-range correlations decay.

\begin{table}[ht]
\begin{center}
\begin{tabular}{lllll}
\hline\hline
    Reference & $\nu$ & $\gamma$ & $\beta$ & $\eta$ \\
\hline\\ Weinrib, Halperin, \cite{Weinrib83} & 1 & 2 & 1/2 & 0
\\ Prudnikov {\em et al.}, \cite{Prudnikov} & 0.7151 & {\it \small 1.4449}
& {\it \small 0.3502} & -0.0205\\ Ballesteros, Parisi,
\cite{Ballesteros99}$^a$ & 1.012(10)& {\it \small 1.980(16)}& {\it \small
0.528(7)}& 0.043(4)\\
 \cite{Ballesteros99}$^b$ &1.005(14)& {\it \small 1.967(23)}&{\it \small
0.524(9)}& 0.043(4)\\ Prudnikov {\em et al.}, \cite{Prudnikov05}$^a$
& 0.719(22) &{\it \small 1.407(24)}& 0.375(45) & {\it \small 0.043(93)}\\
\cite{Prudnikov05}$^b$  & 0.710(10)  & 1.441(15)&0.362(20)  &{\it \small
-0.030(7)}
\\
 \hline
\end{tabular}
\end{center}
\caption{\label{tab1} The critical exponents for the
three-dimensional Ising model with extended impurities for $a=2$.
Renormalization group calculations: first order
$\varepsilon,\delta$-expansion \cite{Weinrib83}; two-loop massive
renormalization scheme \cite{Prudnikov}. Monte Carlo simulations:
finite-size scaling, combination of Wolff and Swendsen-Wang
algorithms, magnetic site concentration $p=0.8$
\cite{Ballesteros99}$^a$, $p=0.65$ \cite{Ballesteros99}$^b$;
short-time critical dynamics with Metropolis algorithm, $p=0.8$
\cite{Prudnikov05}$^a$, finite-size scaling with Wolff algorithm,
$p=0.8$ \cite{Prudnikov05}$^b$. Values of the exponents obtained
from the scaling relations are shown in italic (see section \ref{II}
for a more detailed discussion).}
\end{table}

The outline of the paper is the following. In the next section we
make a brief overview of the results of previous analysis of the 3d
Ising model with long-range-correlated impurities paying special
attention to the former MC simulations. Details of our MC
simulations are explained in sections \ref{III} and \ref{IV}. There,
we formulate the model and define the observables we are interested
in. We analyze statistics of typical and rare events taking the
magnetic susceptibility as an example in section \ref{IV}. The
numerical values of the exponents are evaluated there by the
finite-size scaling technique. Section \ref{V} concludes our study.

\section{Overview of previous analytical and MC results}\label{II}

Currently, there exist two different analytical results for the
values of critical exponents of the 3d Ising model with
long-range-correlated impurities. The first one is due to Weinrib
and Halperin, who formulated the model of a $m$-vector magnet with
quenched impurities correlated via the power law (\ref{1})
\cite{Weinrib83}. They used the renormalzation group technique
carrying out a double expansion in  $\varepsilon=4-d$, $\delta=4-a$,
considering $\varepsilon$ and $\delta$ to be of the same order of
magnitude, and estimating values of the critical exponents to the
first order in this expansion. Further, they conjectured the
obtained first order result for the correlation length critical
exponent
 \begin{equation} \label{3}
 \nu=2/a
 \end{equation}
to be an exact one and to hold for any value of spin component number
$m$ provided that $\nu^{\rm pure}<2/a$. By complementing Eq.~(\ref{3})
with the first order value of the pair correlation function critical
exponent, $\eta=0$ \cite{Weinrib83}, the other critical exponents can
be obtained via familiar scaling relations. For $a=2$ the exponents
are listed in table \ref{tab1}.

The second theoretical estimate was obtained by Prudnikov {\em et al.}
\cite{Prudnikov} in the field theoretical renormalization group
technique by performing renormalization for non-zero mass at fixed
space dimension $d=3$. Their two-loop calculations refined by the
resummation of the series obtained bring about a non-trivial
dependence of the critical exponents both on $m$ and on $a$. We list
their values of the exponents for $m=1$ and $a=2$ in the second row of
table \ref{tab1}. In particular, one observes that the correlation
length exponent $\nu$ obtained in Ref. \cite{Prudnikov} differs from
the value predicted by (\ref{3}) by the order of 25 \%. As for the
reason of such discrepancy the authors of Ref. \cite{Prudnikov} point
to a higher order of the perturbation theory they considered together
with the methods of series summation as well as to taking into account
the graphs which are discarded when the $\varepsilon,\delta$-expansion
is being used.

Let us note however, the qualitative agreement between the above
analytical results: both renormalization group treatments, Refs.
\cite{Weinrib83} and \cite{Prudnikov}, predict that the new
(long-range-correlated) fixed point is stable and reachable for the
condition considered and hence the 3d Ising model with
long-range-correlated impurities belongs to a new universality
class. These are only the numerical values of the exponents which call
for additional verification. In such cases one often appeals to the
numerical simulations. Indeed, two simulations were performed, and,
strangely enough, the results again split into two groups: whereas the
simulation of Ballesteros and Parisi \cite{Ballesteros99} is strongly
in favour of the theoretical results of Weinrib and Halperin, the
simulation of Prudnikov {\it et al.}  \cite{Prudnikov05} almost
exactly reproduces former theoretical results of this group
\cite{Prudnikov} as one can see from table 1. There, we give the
results for the exponents obtained in simulations complementing them
by those that follow if one uses familiar scaling relations
(the last are shown in italic).

Since our own study will rely on the simulation technique as well, we
describe an analysis performed in Refs.
\cite{Ballesteros99,Prudnikov05} in more details.  In Ref.
\cite{Ballesteros99} the simulations were done for two different types
of the long-range-correlated disorder referred by the authors as the
Gaussian and the non-Gaussian one. In the first case, the point-like
defects are scattered directly on the sites of a 3d simple cubic
lattice according to the desired decay of the pair correlation
function (\ref{1}), $a=2$. In the second case, the defects form the
lines directed along the randomly chosen axes and, as was mentioned
before, their impurity-impurity pair correlation function should also
decay at large $r$ as $g(r)\sim 1/r^2$). The MC algorithm used in Ref.
\cite{Ballesteros99} is a combination of a single-cluster Wolff method
and a Swendsen-Wang algorithm. After a fixed number of a
single-cluster updates one Swendsen-Wang sweep is performed. The
procedure was called a MC step (MCS). The number of single cluster
flips was chosen such that the autocorrelation time $\tau_E$ was
typically 1 MCS. For thermalization of the system $100$ MCSs were
performed and then various observables were measured. The simulation
was done on the lattice of sizes $L=8,16,32,64$ and $128$. Then $100$
MCSs have been used for measurements on $N_S$=20000 different samples
for $L\leq 64$ and on $N_S$=10000 samples for $L=128$. The procedure
seems to be quite safe.

Values of the critical exponents $\nu$ and $\eta$ obtained in Ref.
\cite{Ballesteros99} in the case of Gaussian disorder (point-like
long-range-correlated defects) are given in table \ref{tab1} for the
impurities concentrations $p=0.8$, $p=0.65$. It was stated in the
paper, that for the non-Gaussian disorder (lines of defects) the
estimates of the exponents are comparable with analytic calculations
of Weinrib and Halperin \cite{Weinrib83}.

In Ref. \cite{Prudnikov05} the MC simulation of the critical
behavior of the three-dimensional Ising model with
long-range-correlated disorder at criticality was performed by means
of the short-time dynamics and the single-cluster Wolff
method. Randomly distributed defects had a form of lines and
resembled the "non-Gaussian disorder" of Ref. \cite{Ballesteros99}.
However, in contrast to Ref. \cite{Ballesteros99}, a condition of
lines mutual avoidance  was implemented. According to Ref.
\cite{Prudnikov05}, a situation when crossing of the lines of
defects is allowed is not described by the Weinrib-Halperin model.
In the short-time critical dynamics method, the concentration of
spins was chosen $p=0.8$ and the cubic lattices of sizes
$L=16\div128$ were considered. As a MC method was used a Metropolis
algorithm.  Resulting numbers for the exponents are given in the 5th
row of table \ref{tab1}.

Additional simulations in the equilibrium state were performed in Ref.
\cite{Prudnikov05} to verify the reliability of results obtained by
means of the short-time critical dynamics. The single-cluster Wolff
algorithm was used for simulation and a finite size scaling for
evaluation of the critical exponents. One MCS was defined as 5 cluster
flips, $10^4$ MCSs were discarded for equilibration and $10^5$ for
measurement. Disorder averaging was typically performed over $10^4$
samples. Again, the procedure seems to be safe.
The values of critical exponents obtained in the simulations
are quoted in the 6th row of table \ref{tab1}.

With the above information at hand we started our numerical analysis
as explained in the forthcoming sections.

\section{Model and observables} \label{III}

\begin{figure}[!ht]
\centerline{\epsfxsize=7cm \epsffile{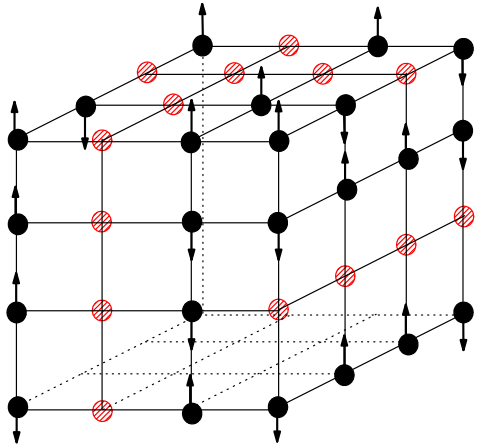}
\epsfxsize=7cm}
\caption{\label{fig_1} A 3d Ising model with long-range-correlated
disorder in the form of randomly oriented lines of non-magnetic
sites. Magnetic sites are shown by discs with arrows, non-magnetic
sites are dashed (red on-line) and do not hold arrows.}
\end{figure}

We consider a $d=3$ Ising model with non-magnetic sites arranged in
a form of randomly oriented lines (see Fig. \ref{fig_1}). The
Hamiltonian reads:
\begin{equation}
{\mathcal H} = -J\sum_{\langle ij\rangle }c_ic_jS_iS_j, \label{Ham}
\end{equation}
where $\langle ij\rangle $ means the nearest neighbour summation over
the sites of a s.c. lattice of linear size $L$, $J>0$ is the
interaction constant, Ising spins $S_i=\pm 1$, $c_i = 0; 1$ is the
occupation number for the $i$-th site and periodic boundary conditions
are employed. Non-magnetic sites ($c_i = 0$) are quenched in a fixed
configuration and form lines, as shown in Fig.
\ref{fig_1}. Concentration $p$ of the magnetic sites is taken to be
far above the percolation threshold. Fig. \ref{fig_1} shows one
possible configuration for non-magnetic impurity lines. During the
simulations, one generates a large number of such disorder
realizations. This is done in our study using the following
procedure. The lines of impurities are generated one by one, each
along randomly chosen axis and the final disorder realization is
accepted with the probability
\begin{equation}\label{probp}
P(p) = \exp(-(p-p_{\rm req})^2/\sigma^2)
\end{equation}
using rejection method. Here $p_{\rm req})$ is a required value
for the concentration of magnetic sites and $\sigma$ is the
dispersion of the resulted Gaussian distribution in $p$ (see
Fig.~\ref{hist}). The average value for the impurity concentration
is equal to $1-p$. In the simulations presented here we paid
special attention to the width of this distribution and consider
both broad and narrow cases, see sections \ref{IVb} and
\ref{IVc}).

The observables saved during each step of the MC simulations for a
given disorder realization will be the instantaneous values for the
internal energy ${\mathcal E}$ and magnetisation ${\mathcal M}$ per
spin, defined as
\begin{equation}
{\mathcal E} = - J\frac{1}{N_p}\sum_{\langle ij\rangle
}c_ic_jS_iS_j, \label{Ene}
\end{equation}
\begin{equation}
{\mathcal M} = \frac{1}{N_p}\sum_{i}c_iS_i, \label{Mag}
\end{equation}
where number of magnetic sites is $N_p=pN$, total number of sites is
$N=L^3$. Using the histogram reweighting technique
\cite{Ferrenberg91} we compute the following expectation values at
the critical temperature region for a given disorder realization:
\begin{equation}\label{99}
{\langle{\mathcal E}\rangle}, \hspace{2em}
{\langle|{\mathcal
M}|\rangle}, \hspace{2em} {\langle{\mathcal M}^2\rangle},
\hspace{2em} {\langle{\mathcal M}^4\rangle}
\end{equation}
as functions of an inverse temperature $\beta=1/T$. In (\ref{99})
the  angular brackets $\langle \dots \rangle$ stand for the
thermodynamical averaging in one sample of disorder.

For a given disorder configuration, one can evaluate the temperature
behaviour of the magnetic susceptibility $\chi$ and magnetic
cumulants $U_2$, $U_4$:
\begin{equation}\label{22a}       \hspace{-1em}
\chi=\beta J N_p (\langle{\mathcal M^2}\rangle-
\langle{|\mathcal M|}\rangle^2), \hspace{1em}  U_4 = 1 - \frac{\langle {\mathcal
M^4}\rangle }{3\langle {\mathcal M^2}\rangle ^2} , \hspace{1em} U_2
= 1 - \frac{\langle {\mathcal M^2}\rangle }{3 \langle{|\mathcal M|}\rangle^2}.
\end{equation}

\begin{figure}[!ht]
\centerline{\epsfxsize=6cm \epsffile{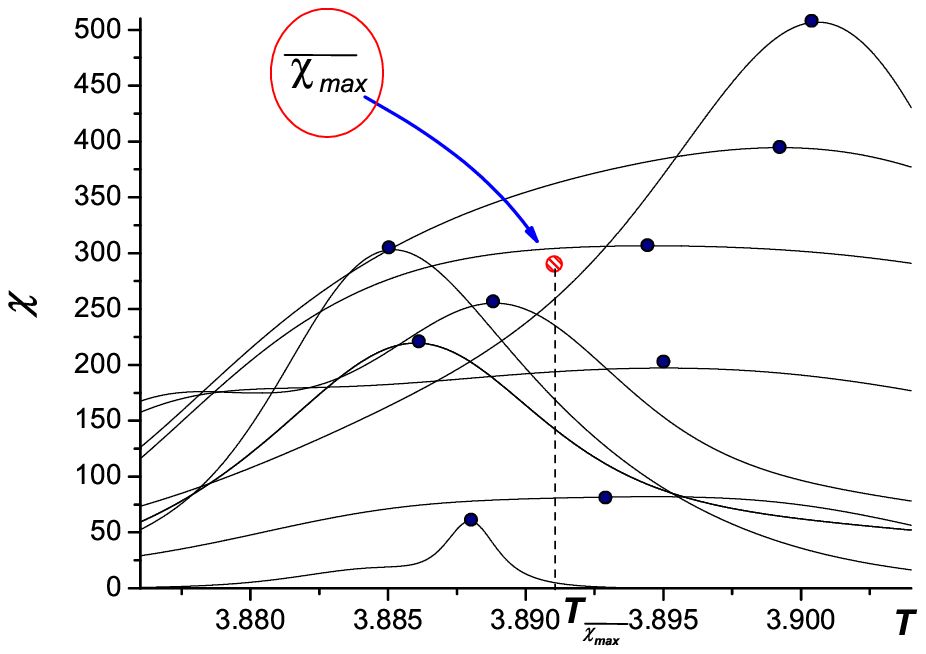} \hspace{3em}
\epsfxsize=6cm \epsffile{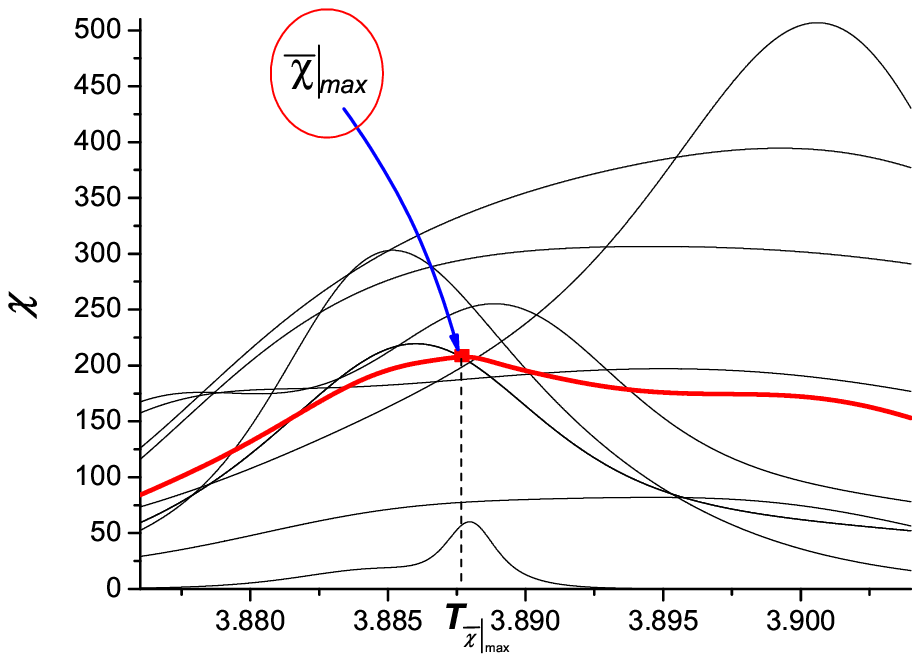}} \centerline{a) {\em Averaging
a} \hspace{9em} b) {\em Averaging b}}
 \caption{\label{fig_11} Two ways to perform an averaging over
 disorder configurations. {\em Averaging a}: the maximal values of
magnetic susceptibility are found for each disorder realization (black discs) and
averaged over disorder configurations. The resulting averaged value
$\overline{\chi_{\rm max}}$ is shown by the red square. {\em Averaging b}: the
configurational averaging of $\chi$ is performed (red dashed curve) and a maximum of
the resulting curve $\overline{\chi}|_{\rm max}$ is found (red square).}
\end{figure}

In order to refer to the physical quantities, the observables are to
be averaged over different disorder configurations, denoted hereafter
by an overline: $\overline{(\dots)}$. Two ways of averaging can be
found in the literature, which we will consider for the case of
finding the maximum value for the susceptibility. At each disorder
realization an individual curve for the susceptibility $\chi$ as a
function of the temperature $T$ is obtained (using histogram
reweighting technique). In the first method of averaging, depicted in
Fig. \ref{fig_11},a the maximal values of each individual curves are
averaged over all disorder realizations. This type of averaging,
referred hereafter as {\em averaging a}, will be denoted as
$\overline{\chi_{\rm max}}$.
An alternative method is shown in Fig. \ref{fig_11},b: at first one
performs a configurational averaging of the susceptibility, i.e. the
single averaged curve is evaluated for $\overline{\chi(T)}$. Then the
maximal value for this curve, $\overline{A}|_{\rm max}$, is
found. Hereafter, this method will be referred as {\em averaging
b}. Both {\em averaging a} and {\em averaging b} will be exploited in
our analysis.

To evaluate the critical exponents, a number of characteristics will
be used with known finite size scaling:
\begin{itemize}
\item {\em averaging a}:
\begin{equation} \label{111}
\overline{\chi_{\rm max}}, \hspace{1em} \overline{{\langle
|{\mathcal M}|\rangle}_{\rm max}}, \hspace{1em}
D_{U_4}=\overline{{\left.\frac{\partial U_4}{\partial
\beta}\right|}_{\rm max}}, \hspace{1em}
D_{U_2}=\overline{{\left.\frac{\partial U_2}{\partial
\beta}\right|}_{\rm max}}.
\end{equation}
Here,
$\overline{\chi_{\rm max}}$, $D_{U_4}$, and $D_{U_2}$ are the averaged over disorder
configurations maximal values of the magnetic susceptibility and magnetic cumulants
temperature derivatives. The value $\overline{{\langle |{\mathcal M}|\rangle}_{\rm max}}$ is
calculated for the temperature that corresponds to the magnetic
susceptibility maximum.

\item {\em averaging b}:

\begin{equation}
\left.\overline{\chi}\right|_{\rm max}, \hspace{1em} \left.
{\overline{\langle |{\mathcal M}|\rangle}}\right|_{\rm max},
\hspace{1em} {\mathcal D}_{U_4}= {\left.\frac{\partial {\mathcal
U}_4}{\partial \beta}\right|}_{\rm max}, \hspace{1em}  {\mathcal
D}_{U_2}={\left.\frac{\partial {\mathcal U}_2}{\partial
\beta}\right|}_{\rm max}, \label{111e} \end{equation}
\begin{equation}
 {\mathcal D}_{M}= {\left.\frac{\partial
\log \overline{\langle|{\mathcal M}|\rangle}}{\partial
\beta}\right|}_{\rm max}, {\mathcal D}_{M^2}={\left.\frac{\partial
\log \overline{\langle{\mathcal M}^2\rangle}}{\partial
\beta}\right|}_{\rm max} , \label{111a}
\end{equation}
where
\begin{equation}\label{22b}
{\mathcal U}_4 = 1 - \frac{\overline{\langle {\mathcal M^4}\rangle}
}{3{\overline{\langle {\mathcal M^2}\rangle }}^2}, \hspace{2em}
{\mathcal U}_2 = 1 - \frac{\overline{\langle {\mathcal M^2}\rangle}
}{3{\overline{\langle {| \mathcal M |}\rangle }}^2}.
\end{equation}
\end{itemize}
Note, that quantities (\ref{111a})  are ill-defined if {\em averaging a}
is considered. Moreover, as one can see from the definitions
(\ref{22a}), (\ref{22b}) there are different ways to define magnetic
cumulants for the disordered system (see Refs. \cite{Janke} for more
discussion).

We performed our numerical simulations using the Swendsen-Wang cluster
algorithm. The main reason is, that with the diluted system one often
performs simulations at the temperature which is far from the native
phase transition point of each particular disorder realization. As the
result, the single clusters can be of rather small size and the Wolff
one algorithm, used by us before \cite{Ivaneyko06a} is found to be
less effective.  The following set of lattice sizes $L=6\div96$
($L=6\div128$ in some cases) is used and performing the
finite-size-scaling analysis is performed.  The magnetic site
concentration is chosen to be $p=0.8$ which is far from the
percolation threshold and this also allows comparison with the
previous simulations \cite{Ballesteros99,Prudnikov05}. Another reason
for such a choice is that for the 3d Ising model with uncorrelated
impurities the correction to scaling terms were found to be minimal at
$p=0.8$ \cite{reviews,Ballesteros98}, (obviously, this does not mean
that $p=0.8$ minimizes correction-to-scaling terms at any level of
impurities correlation). The number of samples ranges from $10^3$  to $10^4$ for
all lattice sizes. The simulations are performed at the finite-size
critical temperature $T_c(L)$ which is found from the maximum of
magnetic susceptibility $\chi$ at each lattice size $L$ and the
reweighting technique for the neighbouring temperatures is used. Note,
that in Refs. \cite{Ballesteros99,Prudnikov05} all simulation were
performed at the critical temperature of an infinite system.

Similarly to our other previous papers
\cite{Ivaneyko05,Ivaneyko06} we start simulation from estimating
the critical temperature of a finite-size system as a function of
the lattice size $L$, $T_c(L)$. For the smallest lattice size
$L=6$, the preliminary simulation is performed at $pT_c^{\rm
pure}$ first ($T^{\rm pure}_c$ is the critical temperature for the
pure system). The $T_c(L=6)$ is located then from the maximum of
the susceptibility. At the same time, we estimate various
autocorrelation times to control the error due to time
correlations. For the next larger lattice size, e.g. $L=8$,
preliminary simulation is performed at $T_c(L=6)$ and then
$T_c(L=8)$ is located again from the maximum of the susceptibility
at $L=8$.  The process is repeated for all lattice sizes.  In this
way we estimate $T_c(L)$ and the energy autocorrelation time
$\tau_{{\langle{\mathcal E}\rangle}}$. The results are given in
table \ref{tab2}. One can see that the finite size system critical
temperature obtained via {\em averaging a} differs from those
obtained via and {\em averaging b}. The difference is not too
large, the next simulations were performed at the critical
temperature obtained by {\em averaging a}.
\begin{table}[!h]
\small{
\begin{center}
\begin{tabular}{l|c|c|c}
\hline\hline
 \multirow{2}{*}{L}& \multirow{2}{*}{$\tau_{{\langle{\mathcal E}\rangle}}$}&\multicolumn{2}{|c}{$T_c(L)$} \\
\cline{3-4}
    &        & $\overline{\chi_{\rm max}}$      & $\left.\overline{\chi}\right|_{\rm max}$\\
\hline
6  &2.50051&3.796779&3.795500\\
8  &3.10875&3.812864&3.810200 \\
10 &3.39280&3.821152&3.817200\\
12 &3.58191&3.829904&3.823000\\
16 &3.90468&3.849101&3.842400\\
20 &4.10689&3.855457&3.849200\\
26 &4.28566&3.863750&3.860100\\
35 &4.38278&3.871662&3.869700\\
48 &4.4438 &3.876678&3.874100\\
64 &4.46380&3.881388&3.876900\\
96 &4.26933&3.884435&3.881200\\
128&4.21278&3.886679&3.883220\\
   &       &$3.89124\pm0.00057$&$3.89214\pm0.00190$\\
   &       &$3.89108\pm0.00137$&$3.89033\pm0.00139$\\
 \hline
\end{tabular}
\end{center}
 \caption{\small The critical temperature $T_c(L)$ of finite size
system, obtained from maximum of magnetic susceptibility and
autocorrelation times of absolute value of $\langle{\mathcal
E}\rangle$ in the Swendsen-Wang cluster sweep. \label{tab2}}}
\end{table}

In evaluation of the critical indices we followed the standard
finite-size-scaling scheme as described e. g. in Ref.
\cite{Ferrenberg91}. The estimates for  the correlation length,
magnetic susceptibility, and spontaneous magnetization critical
exponents $\nu$, $\gamma$ and $\beta$ can be obtained from the FSS
behaviour of the following quantities:
\begin{itemize}
\item {\em averaging a}:
\begin{equation} \label{111b}
D_{U_4},\, D_{U_2} \sim L^{1/\nu}, \hspace{2em} \overline{\chi_{\rm
max}} \sim L^{\gamma/\nu}, \hspace{2em} \overline{{\langle
|{\mathcal M}|\rangle}_{\rm max}} \sim L^{-\beta/\nu},
\end{equation}
\item {\em averaging b}:
\begin{equation} \label{111c}
{\mathcal D}_{U_4},\, {\mathcal D}_{U_2}, {\mathcal D}_{M},
{\mathcal D}_{M^2} \sim L^{1/\nu}, \;
\left.\overline{\chi}\right|_{\rm max} \sim L^{\gamma/\nu}, \;
\left.{\overline{\langle |{\mathcal M}|\rangle}}\right|_{\rm max}
\sim L^{-\beta/\nu}.
\end{equation}
\end{itemize}

In Eqs.~(\ref{111b}), (\ref{111c}), the correction-to-scaling terms
have been omitted. Taking them into account the FSS behaviour of the
quantity $A$ attains the form:
\begin{equation}\label{111d}
A \sim L^{\phi}(1+ \Gamma_A L^{-\omega}),
\end{equation}
where $\phi$ is the leading exponent, $\omega$ is the
correction-to-scaling exponent, and $\Gamma_A$ is a
correction-to-scaling amplitude. Hereafter, we perform the
finite-size-scaling analysis in several ways. First, assuming the
correction-to-scaling terms to be small we will fit the data for the
observables to the leading power laws (\ref{111b}), (\ref{111c})
only. Besides that, we will also perform the fits with
correction-to-scaling terms (\ref{111d}) as explained below.

\section{Statistics of the observables. Preliminary estimates of the
critical exponents} \label{IV}

In this section, applying the above described formalism, we give
preliminary estimates of the critical exponents. As we will see,
the estimates for the same exponent obtained on the base of FSS of
different quantities  differ from each other. Moreover, similar
differences arise when the results obtained via {\em averaging a}
and {\em averaging b} are compared. We will discuss possible
reasons for such behaviour and will analyze them further in the
forthcoming section.

The starting point of the analysis performed in this section is
connected with the fact that at the same concentration of
impurities, different samples vary configurationally (geometrically)
as well as in their thermodynamical characteristics. In systems with
uncorrelated dilution these variations are also present (see e.g.
Ref.~\cite{Chatelain05}) although, as we will see below, these are
less pronounced than in the case of the long-range-correlated
disorder. Moreover, although the mean concentration of magnetic
sites is well-defined for an ensemble  samples (and chosen to be
$p=0.8$), the concentration of magnetic sites in each sample vary
from the mean one. Such deviations are caused by possible
intersections of impurity lines in different samples, besides  not
for all lattice sizes one can reach given concentration of
impurities by an integer number of lines. Therefore, to start an
analysis one has to make a choice of a set of sample specifying
allowed values of the concentration dispersion. We start an analysis
by dealing with the set of samples that are characterized by
dispersion of concentrations $\sigma^2=10^{-4}$.

\subsection{Typical, average, and rare values}\label{IVa}

\begin{figure}[!ht]
\epsfxsize=4.5cm \epsffile{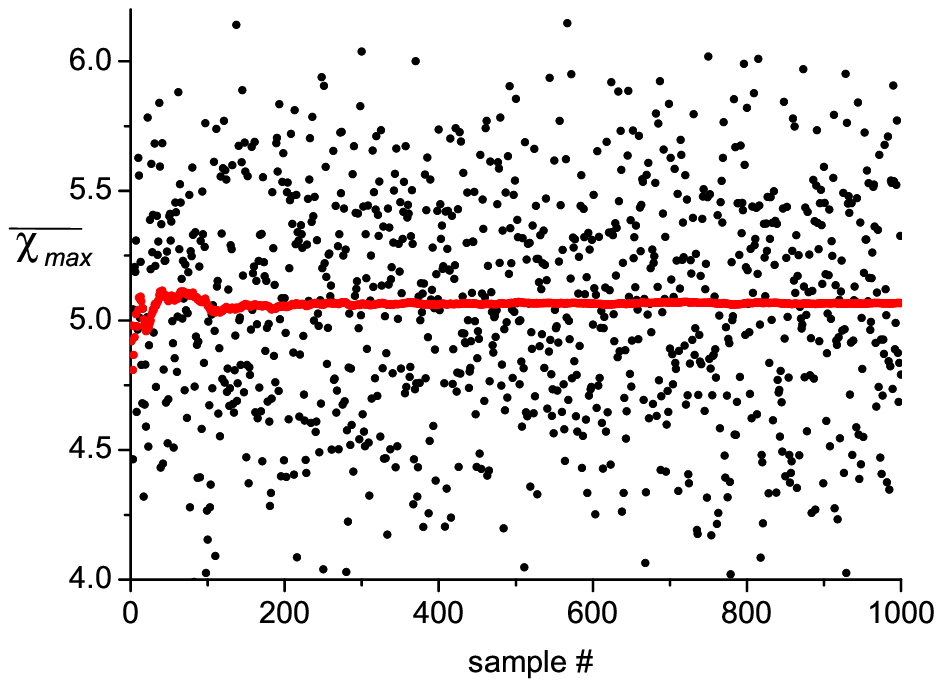} \epsfxsize=4.5cm
\epsffile{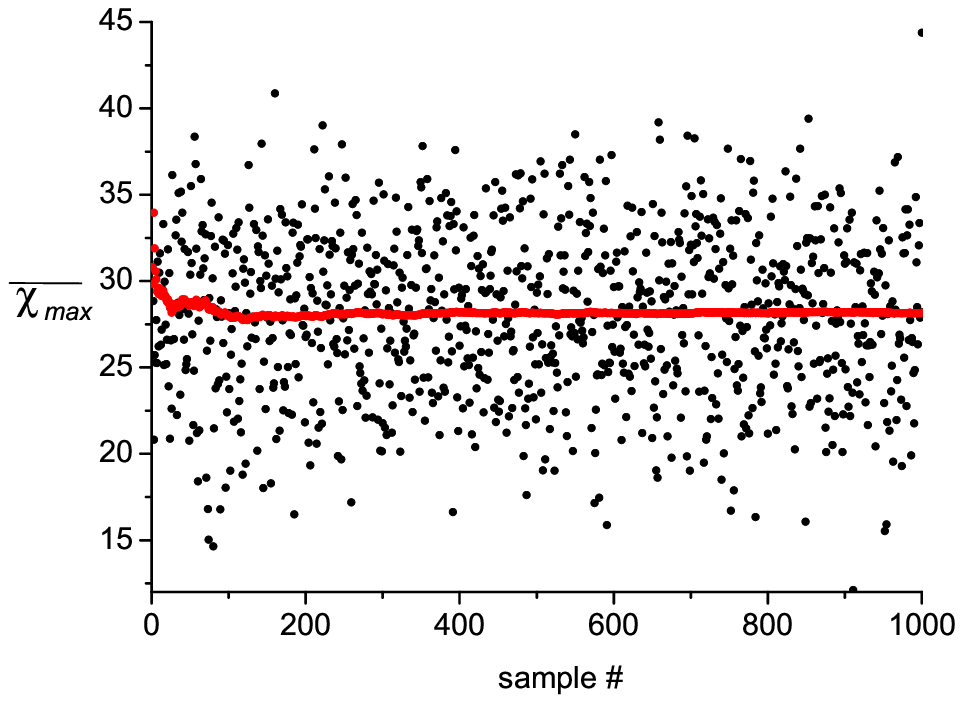}\epsfxsize=4.5cm \epsffile{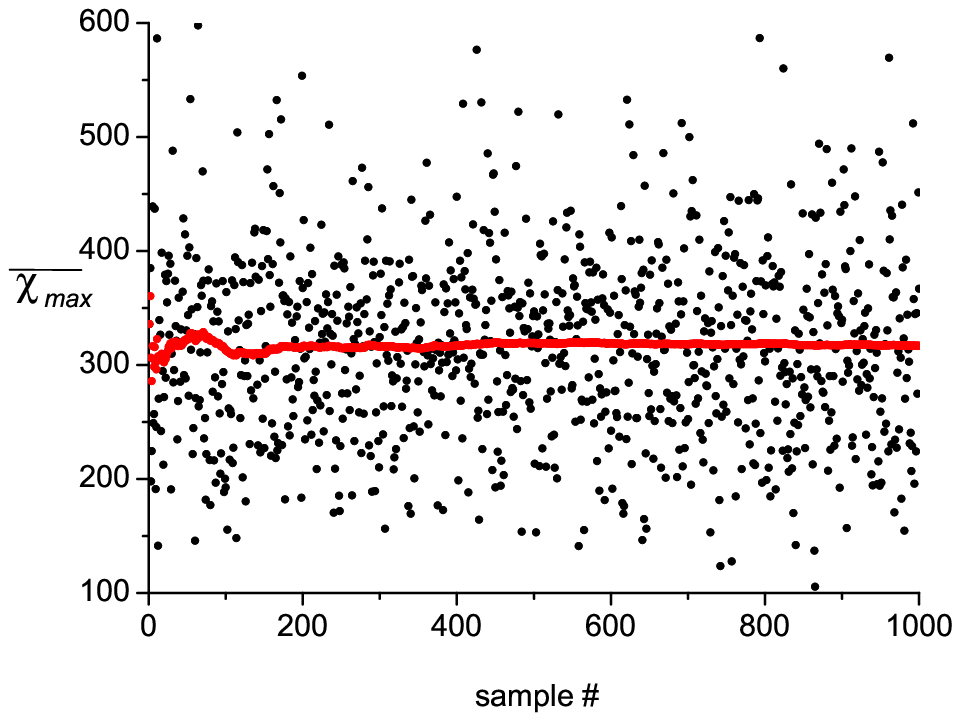}\\
\centerline{(a)\hspace{6cm}(b)\hspace{6cm}(c)} \epsfxsize=4.5cm
\epsffile{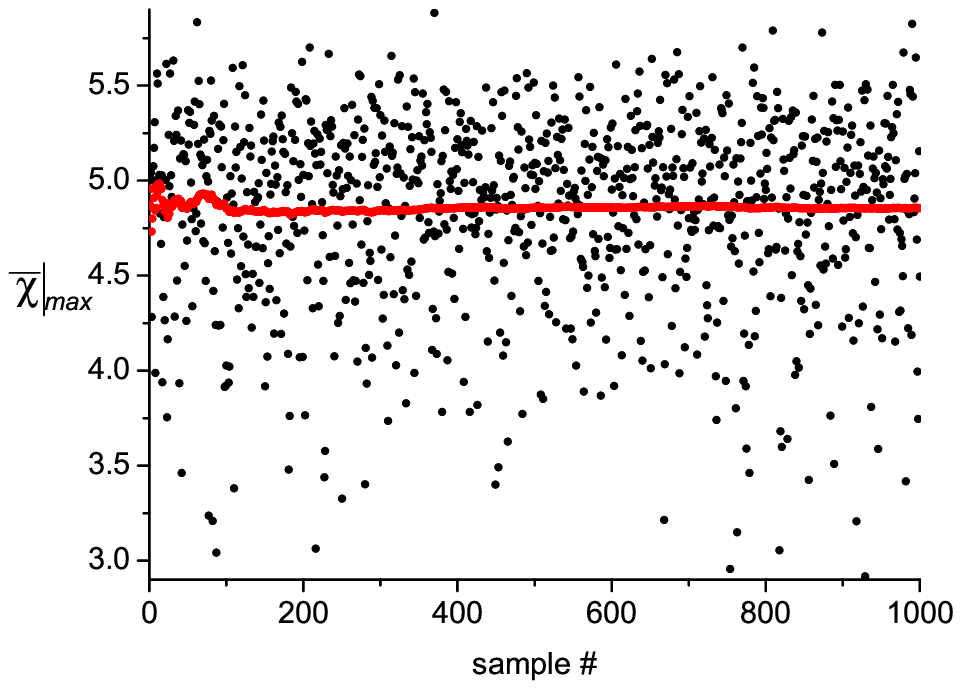} \epsfxsize=4.5cm
\epsffile{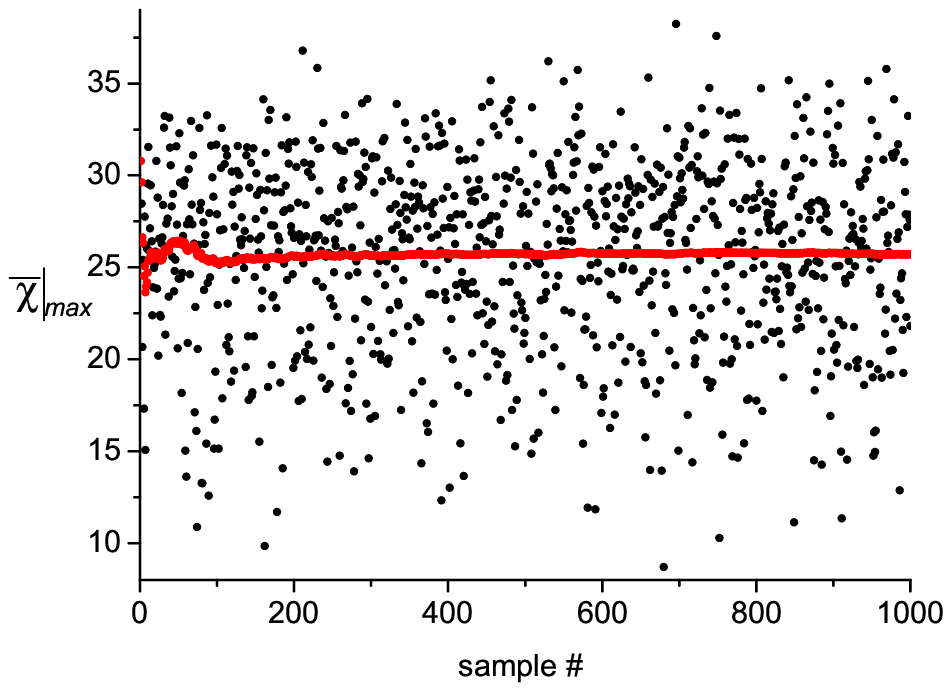}\epsfxsize=4.5cm \epsffile{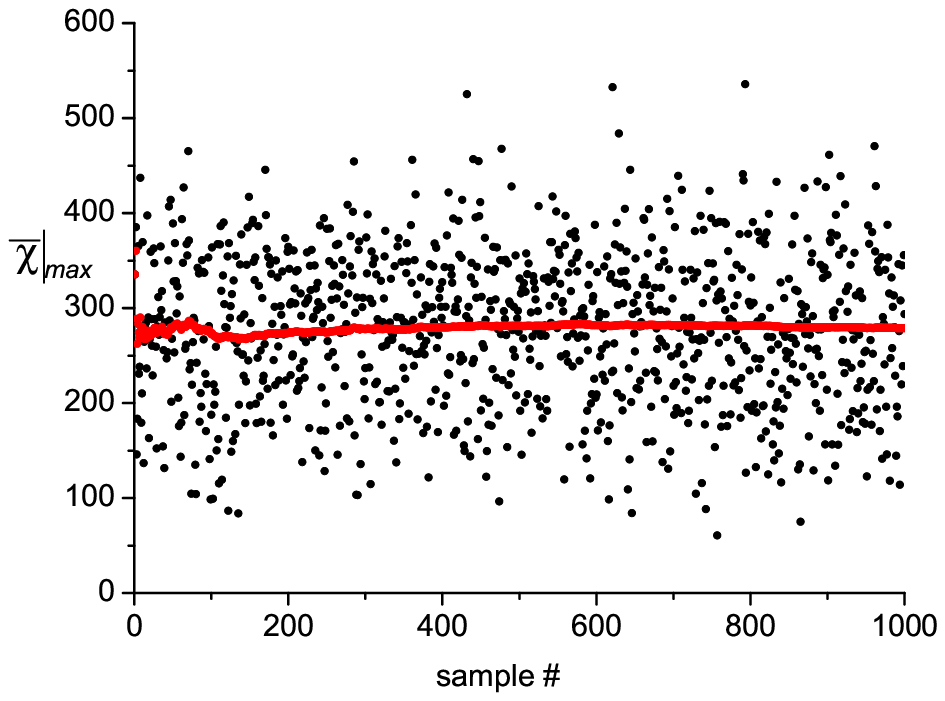}\\
\centerline{(d)\hspace{6cm}(e)\hspace{6cm}(f)}
\caption{\label{fig2} Values of the susceptibility
$\overline{\chi_{\rm max}}$, calculated via {\em averaging a} (
Fig. 3a-3c) and $\left.\overline{\chi}\right|_{\rm max} $,
calculated via {\em averaging b} ( Fig. 3d-3f)  for different
disorder realizations (different samples). Running average over
the samples is shown by the solid line. From left to right: $L =
10,26,96$.}
\end{figure}

The sample-to-sample variations increase with the increase of the
lattice size.  As an example, we show in Fig. \ref{fig2} the values
of the susceptibility for a given disorder realization calculated
via {\em averaging a} ($\overline{\chi_{\rm max}}$, Fig. 3a-3c) and
{\em averaging b} ($\left.\overline{\chi}\right|_{\rm max} $, Fig.
3d-3f) for three typical lattice sizes $L=10,26,96$. Each point in
these plots represents the data obtained for a given sample (i.e.
given disorder realization). Solid lines are made up of the average
values, where at each $\sharp$ the averaging is done on the interval
$0-\sharp$. One can see that the averages saturate at $\sharp$
larger than two hundred samples. From this in particular one can
conclude that considered here  number of samples $N_{dis}=1000$ is
quite sufficient for further analysis.

One observes that for some samples $\overline{\chi_{\rm max}}$ and
$\left.\overline{\chi}\right|_{\rm max} $ differ significantly
from the average values. When this is the case, we will call such
an event a rare one.  The values that are close to the maximal one
will be referred to as typical events. We would like to stress
that for {\em averaging b} for the lattice size $L=10$ most of
rare events produce the values for the susceptibility below the
average value, while for $L=96$ the situation is reversed, see
figure \ref{fig2}. At $L=26$ the distribution of rare events is
aproximately symmetric. The situation is more homogenous  for the
{\em averaging a}.

\begin{figure}[!ht]
\epsfxsize=4.5cm \epsffile{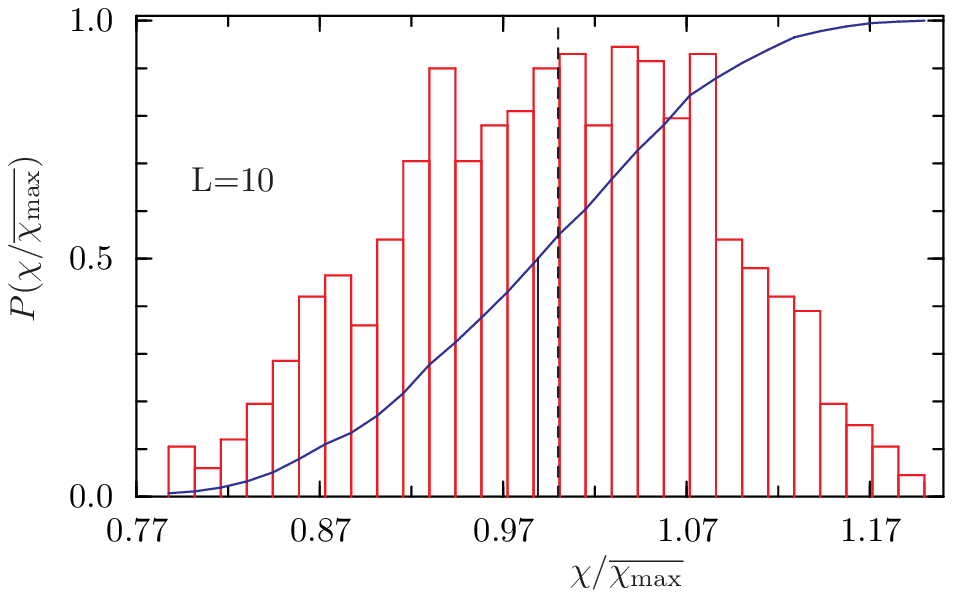}\epsfxsize=4.5cm
\epsffile{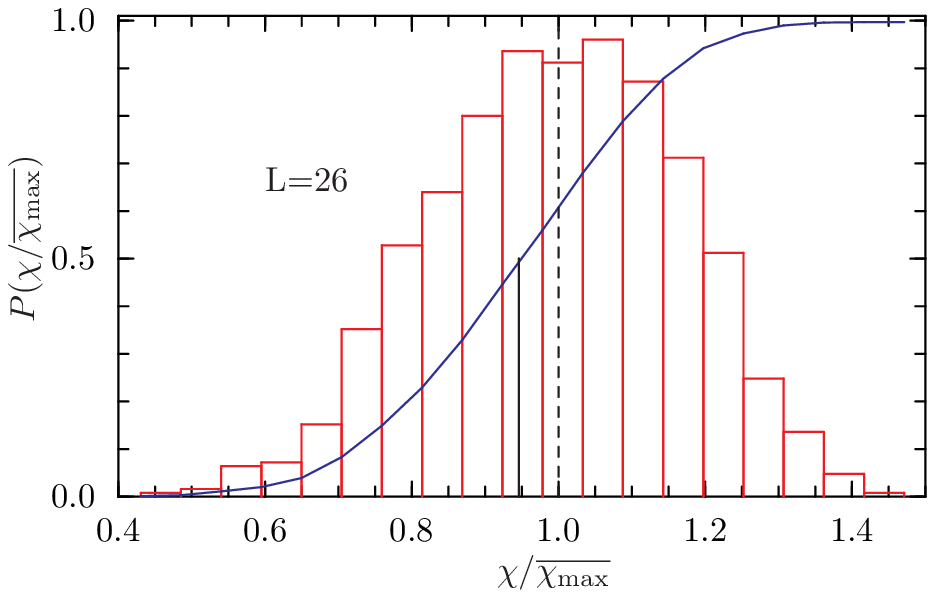}\epsfxsize=4.5cm \epsffile{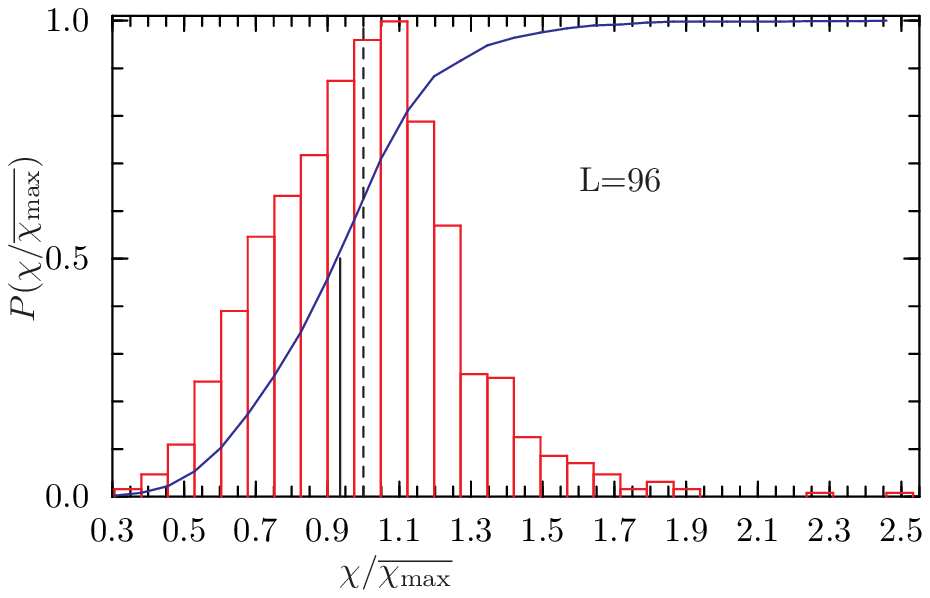}\\
\centerline{(a)\hspace{6cm}(b)\hspace{6cm}(c)} \epsfxsize=4.5cm
\epsffile{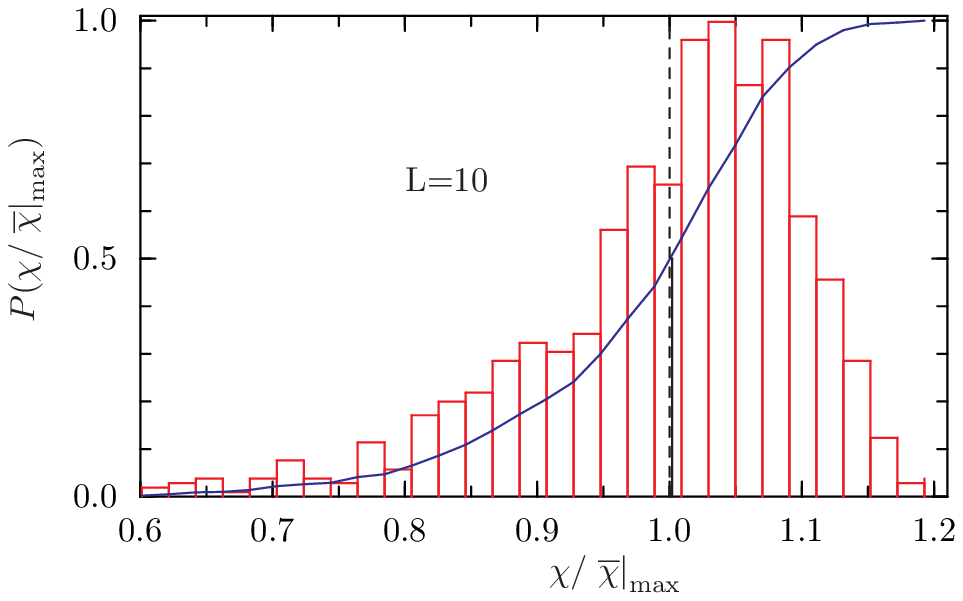}\epsfxsize=4.5cm
\epsffile{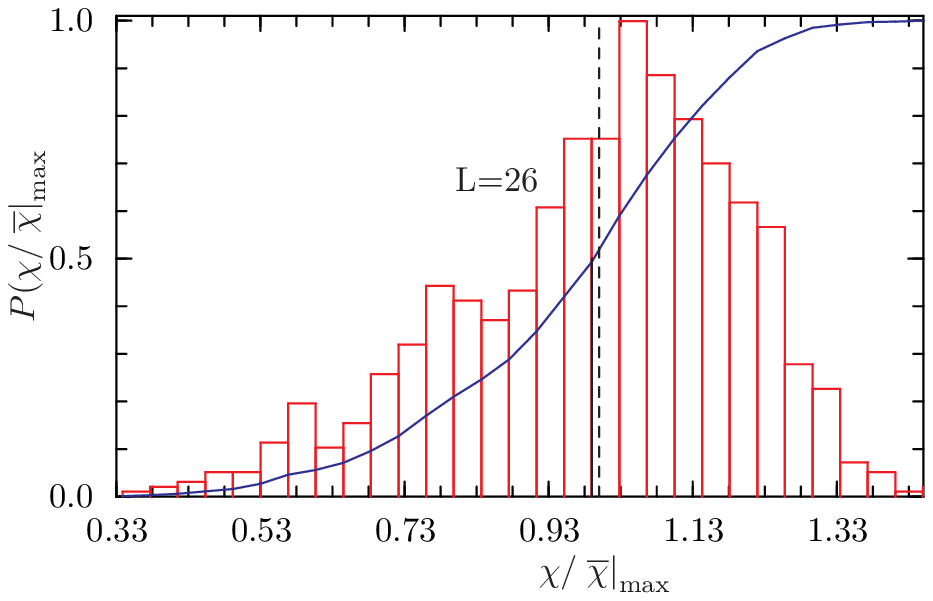}\epsfxsize=4.5cm \epsffile{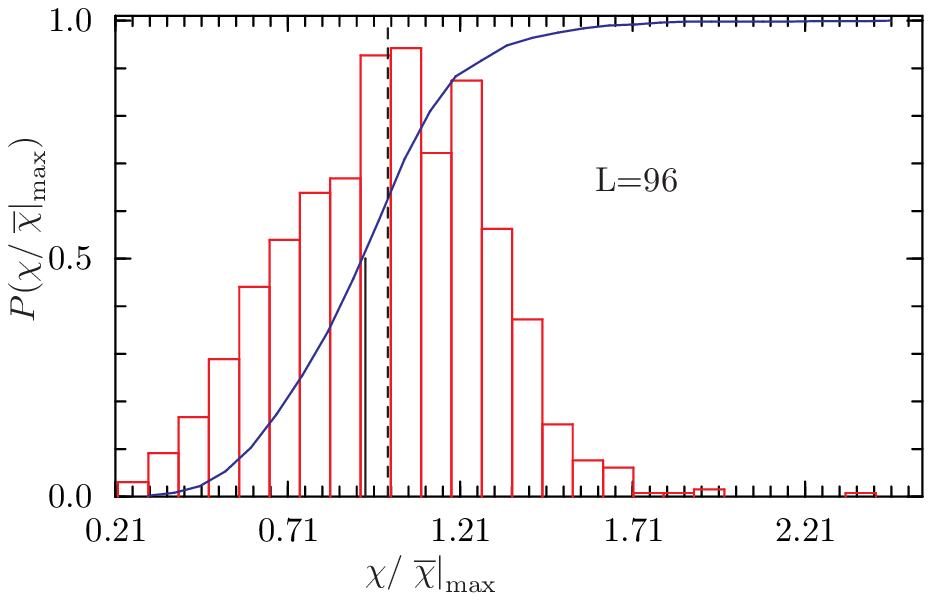}\\
\centerline{(d)\hspace{6cm}(e)\hspace{6cm}(f)}
\caption{\label{fig3}Probability distributions of the
susceptibilities $\overline{\chi_{\rm max}}$,  (Fig. 4a-4c) and
$\left.\overline{\chi}\right|_{\rm max} $, (Fig. 4d-4f) for the
lattice sizes $L=10,26,96$ (from left to right). The full curves
(blue on-line) represent the integrated distribution. Dashed
vertical lines (blue on-line) show $\overline{\chi_{\rm max}}$,
and $\left.\overline{\chi}\right|_{\rm max} $, calculated with the
complete probability distribution. Solid lines  show the median.}
\end{figure}

The averaging over too small number of disorder realizations leads
to typical (i.e. most probable) values instead of averaged ones.
Indeed, as can be seen in Fig. \ref{fig3} for $L=96$, the
probability distributions of $\overline{\chi_{\rm max}}$ and
$\left.\overline{\chi}\right|_{\rm max} $ possess long tails of
rare events with larger values of susceptibility.  The samples
which correspond to the rare events give a large contribution to
the average and shift it far from the most probable values.  The
value of $\left.\overline{\chi}\right|_{\rm max} $ calculated with the complete
probability distribution in shown in fig. \ref{fig3} by a dashed
line (blue on-line).  Another characteristic value shown in fig.
\ref{fig3} by lines (solid) are the median values $\chi_{med}$,
defined as the values of $\chi$ where the integrated probability
distributions are equal to 0.5. A distance between the average and
median values may serve to quantify the asymmetricity of the
probability distribution.

\begin{figure}[!ht]
\epsfxsize=4.5cm \epsffile{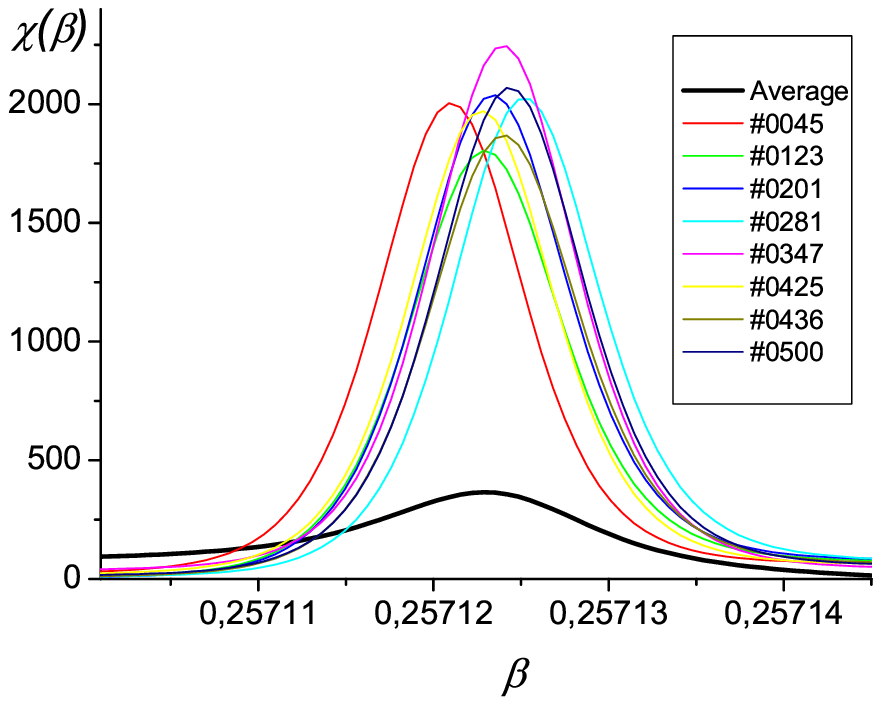}\epsfxsize=4.5cm
\epsffile{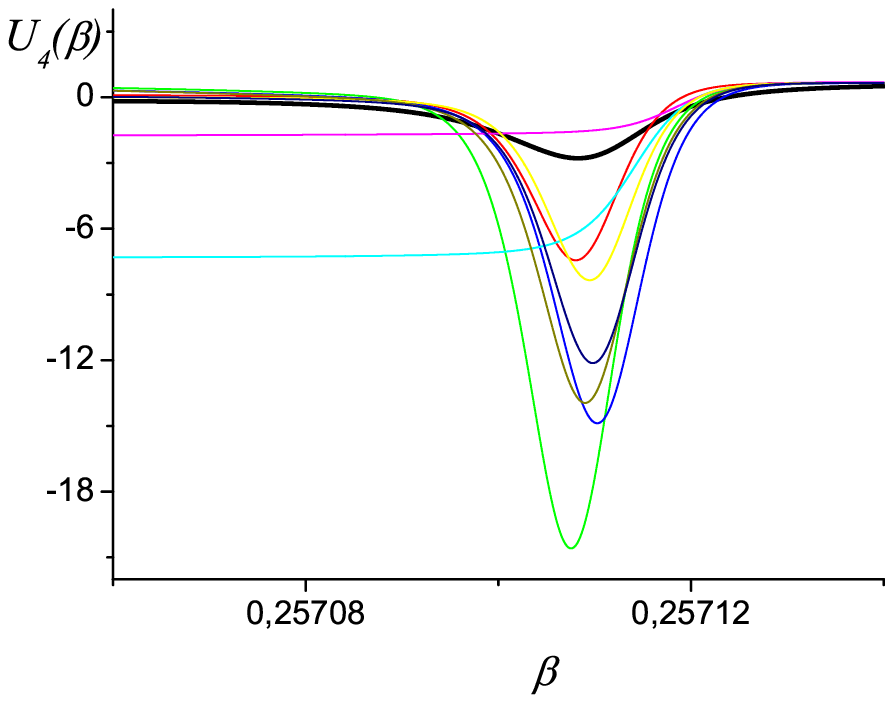} \epsfxsize=4.5cm
\epsffile{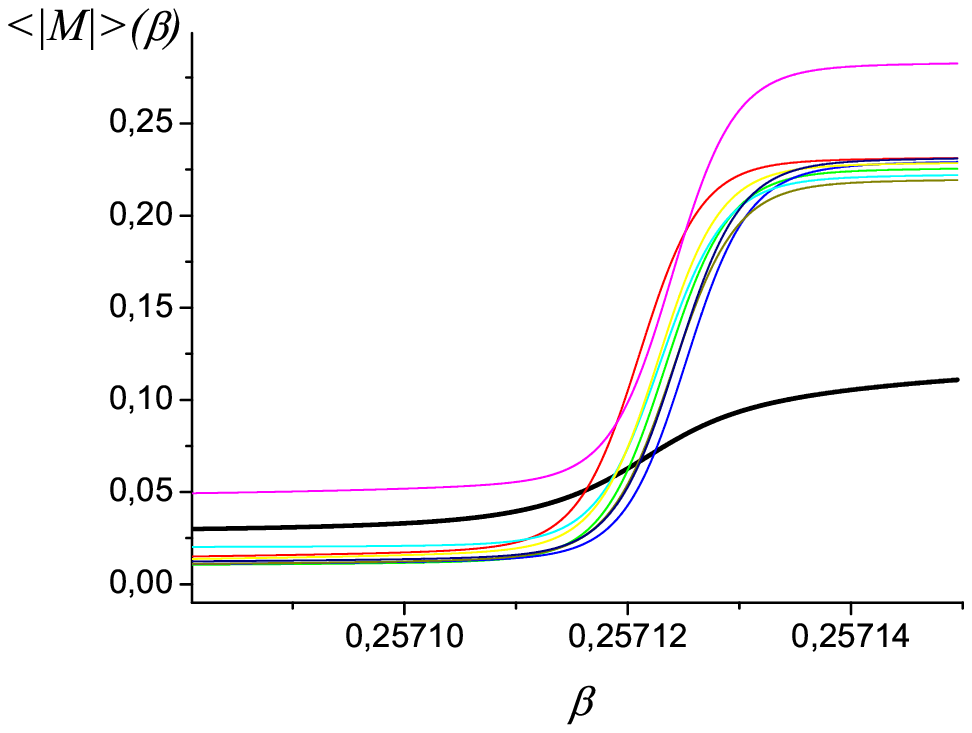} \\
\centerline{(a)\hspace{6cm}(b)\hspace{6cm}(c)}
\caption{\label{fig5a} Examples of {\em typical} events for the
largest lattice size $L=96$ for magnetic susceptibility $\chi$ (a),
Binder's cummulant $U_4$ (b) and magnetisation $\langle|{\mathcal
M}|\rangle$ (c).  Different colours correspond to different samples.
The thick lines show the averages (calculated via {\em averaging b)}
over $10^3$ samples.}
\end{figure}

\begin{figure}[!ht]
\epsfxsize=4.5cm \epsffile{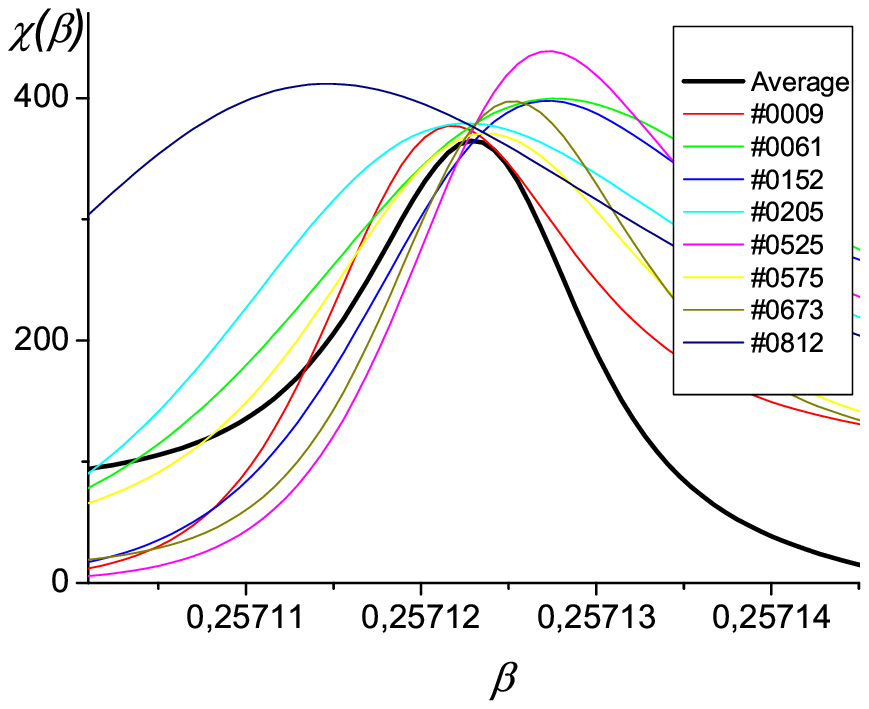}\epsfxsize=4.5cm
\epsffile{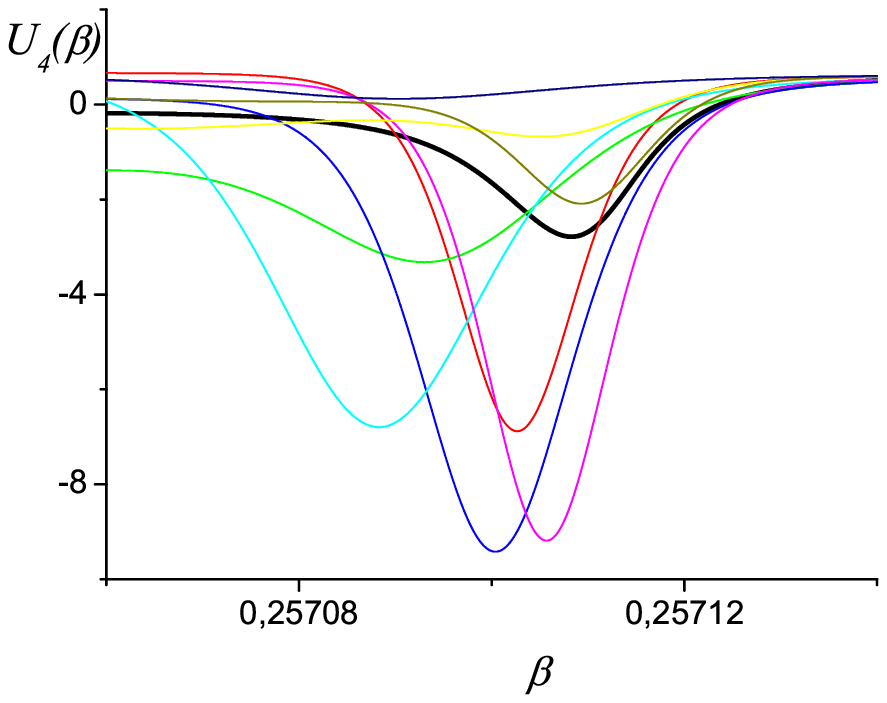} \epsfxsize=4.5cm
\epsffile{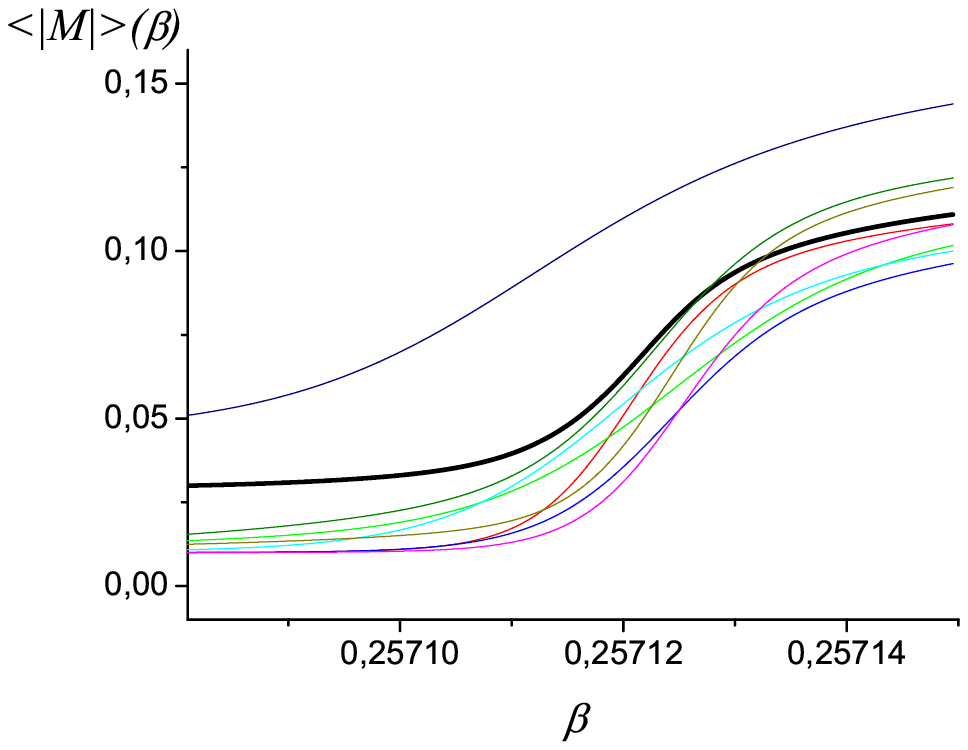}\\
\centerline{(a)\hspace{6cm}(b)\hspace{6cm}(c)}
\caption{\label{fig6a} Examples of {\em rare} events for the largest
lattice size $L=96$ for magnetic susceptibility $\chi$ (a), Binder's
cummulant $U_4$ (b) and magnetisation $\langle|{\mathcal M}|\rangle$
(c). Different colours correspond to different samples. The thick
lines show the averages over all $10^3$~samples.}
\end{figure}

Figures \ref{fig5a} and \ref{fig6a} show contributions of typical
and rare events to the temperature behaviour of the observables
under discussion.  As an example of typical events, we plot in
fig. \ref{fig5a} the magnetic susceptibility $\chi$, Binder's
cummulant $U_4$, and magnetisation $\langle|{\mathcal M}|\rangle$
selecting the samples where susceptibilities have the same values
as $\overline{\chi}$ at $\beta_{max}$ for the largest lattice size
$L=96$. Few examples of rare events corresponding to large values
of $\chi$ are shown in Fig. \ref{fig6a}. The bold lines in the
figures show averages over all disorder realizations (all
samples). The thin lines (color on-line) correspond to different
samples, as shown in the legends to the plots \ref{fig5a}a,
\ref{fig6a}a.

It follows from the above analysis that the rare events are
obviously present in the random variables distributions and give a
contribution to the average values of physical
quantities.\footnote{The use of the familiar $3\sigma$ law to choose
the samples during the MC experiment (as it was done at the initial
phase of our analysis, see Ref. \cite{Ivaneyko06a}) modifies the
distribution. Such modifying leads to different results for the
exponents. We thank the referee for the comments on this point.}

These are the  average values that correspond to the observables and
are to be used for for the FSS analysis. We will perform such
analysis in the next subsection.

\subsection{Evaluation of the critical exponents. Grand canonical disorder}\label{IVb}

Before passing to evaluation of the critical exponents let us recall
that as it was noted at the beginning of section \ref{V} we perform
the simulations for the samples with varying concentration of magnetic
sites, keeping the mean concentration equal to $p=0.8$ and limiting
dispersion of concentrations by $\sigma^2=10^{-4}$ (\ref{probp}).
Following Refs. \cite{Wiseman98} it is natural to call such situation
the 'grand canonical disorder'. Afterwards, in section \ref{IVc} we
will consider the case of 'canonical disorder', when we will choose
the samples much more close in the concentration of magnetic sites
(taking $\sigma^2=10^{-7}$).

\subsubsection{Averaging a}\label{IVb1}

\begin{figure}[!ht]
\epsfxsize=7cm \centerline{\epsffile{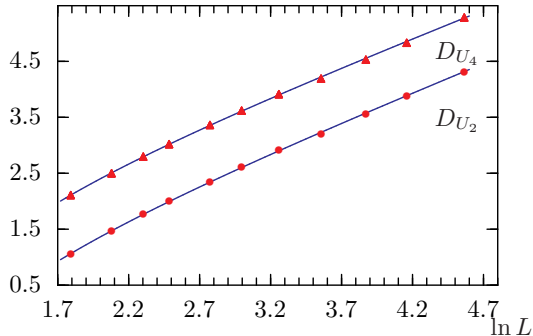}}
\caption{\label{fig5} Log-log plots for the maximum values of the
configurationally averaged derivatives of Binder cumulants
$D_{U_2}$ (discs) and $D_{U_4}$ (triangles). Solid line: data fit
to the power law with the correction-to-scaling, Eq. (\ref{111d}).
Here and below the range of the confidence interval is smaller
than the symbol size.}
\end{figure}

\begin{figure}[!ht]
\epsfxsize=7cm \epsffile{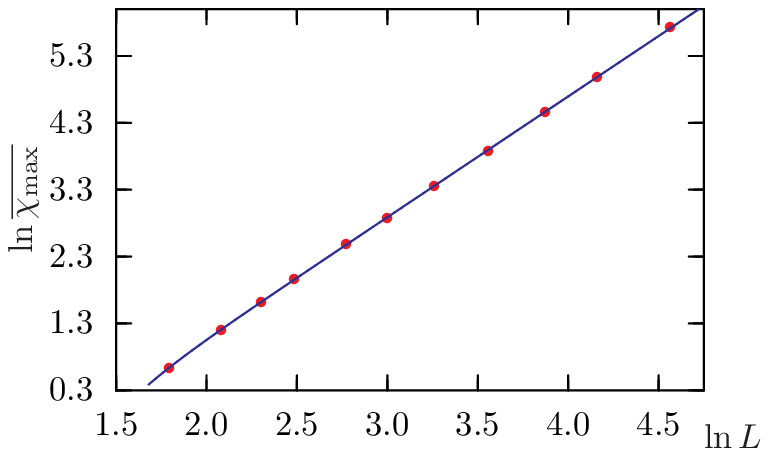} \epsfxsize=7cm
\epsffile{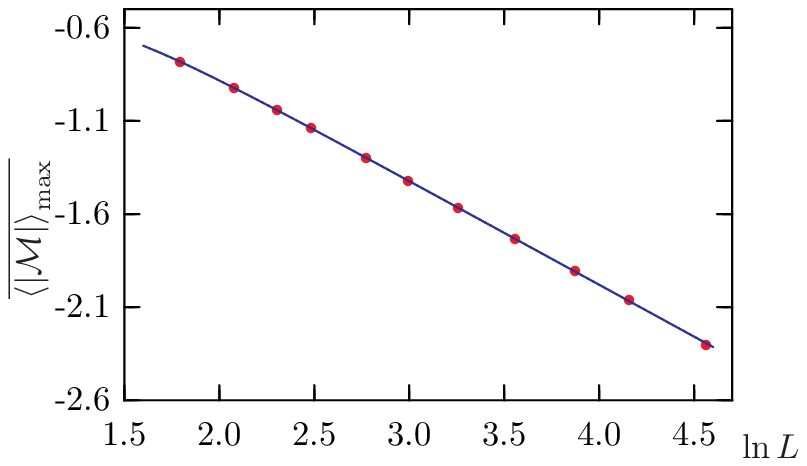} \centerline{(a)\hspace{8cm}(b) }
\caption{\label{fig7} Log-log plots for the configurationally
averaged maximum values of the magnetic susceptibility
$\overline{\chi_{\rm max}}$ and magnetisation
$\overline{{\langle|{\mathcal M}|\rangle}_{\rm max}}$. Solid line:
data fit to the power law with the correction-to-scaling, Eq.
(\ref{111d}).}
\end{figure}

In the finite-size-scaling technique \cite{Ferrenberg91}, the
critical exponents are calculated by fitting the data set for the
observables evaluated at various lattice sizes to the power laws
(\ref{111b}), (\ref{111d}) as was already mentioned in section
\ref{III}. This procedure will be used for the evaluation of the
critical exponents $\nu$, $\beta$, and $\gamma$ based on the
observables measured during the MC simulations. Corresponding
finite-size-scaling plots are shown in Figs. \ref{fig5},
\ref{fig7}, the numbers are given in table \ref{tab3}. In Fig.
\ref{fig5}, we give the log-log plots of the maximum values of the
configurationally averaged derivatives of Binder cumulants
$D_{U_4} $ and $ D_{U_2}$. The scaling of these quantities is
governed by the correlation length critical exponent $\nu$, see
Eqs. (\ref{111b}). Fig. \ref{fig7} shows the size dependence of
the configurationally averaged magnetic susceptibility
$\overline{\chi_{\rm max}}$ (Fig. \ref{fig7}a) and magnetisation
$\overline{{\langle |{\mathcal M}|\rangle}_{\rm max}}$ (Fig.
\ref{fig7}b). These values are expected to manifest a power-law
scaling with exponents $\beta/\nu$ and $\gamma/\nu$, see Eq.
(\ref{111b}).

 \begin{sidewaystable}
\centering
\begin{tabular}{lllll}
\hline\hline & $\overline{\chi_{\rm max}}$&
$\overline{{\langle|{\mathcal M}|\rangle}_{\rm max}}$ &
$D_{U_4}$&$D_{U_2}$
\\
\hline
6  &  $ 1.886\pm0.003$   & $ 0.4584\pm0.0002 $&  $ 8.248\pm0.007 $  &  $2.900\pm0.003$    \\
8  &  $ 3.312\pm0.007$   & $ 0.3976\pm0.0003 $&  $ 12.18\pm0.02  $  &  $4.320\pm0.007$    \\
10 &  $ 5.068\pm0.013$   & $ 0.3533\pm0.0003 $&  $ 16.41\pm0.04  $  &  $5.863\pm0.014$    \\
12 &  $ 7.121\pm0.023$   & $ 0.3205\pm0.0003 $&  $ 20.50\pm0.06  $  &  $7.356\pm0.024$    \\
16 &  $ 11.95\pm0.05 $   & $ 0.2735\pm0.0004 $&  $ 28.87\pm0.13  $  &  $10.45\pm0.05 $    \\
20 &  $ 17.85\pm0.08 $   & $ 0.2413\pm0.0004 $&  $ 37.33\pm0.23  $  &  $13.65\pm0.08 $    \\
26 &  $ 28.66\pm0.14 $   & $ 0.2096\pm0.0004 $&  $ 50.10\pm0.49  $  &  $18.46\pm0.15 $    \\
35 &  $ 48.07\pm0.28 $   & $ 0.1778\pm0.0005 $&  $ 66.14\pm0.71  $  &  $24.58\pm0.25 $    \\
48 &  $ 86.34\pm0.55 $   & $ 0.1491\pm0.0005 $&  $ 92.83\pm1.05  $  &  $34.92\pm0.38 $    \\
64 &  $ 147.3\pm1.0  $   & $ 0.1272\pm0.0005 $&  $ 125.4\pm1.9   $  &  $48.45\pm0.72 $    \\
96 &  $ 310.5\pm2.6  $   & $ 0.1001\pm0.0004 $&  $ 197.1\pm4.5   $  &  $74.71\pm1.49 $  \\
\hline
\end{tabular}
 \caption{\label{tab3} MC simulation data obtained for 3d Ising model with
 long-range correlated disorder: {\em averaging  a}.}
\end{sidewaystable}

We used several ways to extract the values of the exponents from
the data given in Table \ref{tab3} and plotted in Figs.
\ref{fig5}, \ref{fig7}, an outcome is summarized in the table
\ref{tab4}.

\begin{table}[htb]
\begin{center}
\begin{tabular}{l|l|l|l|l}
\hline\hline \multirow{4}{*}{Exponents}&
\multirow{4}{*}{Observables }& \multicolumn{3}{|c}{Data fit
(Number of
data points $n$}\\
& & \multicolumn{3}{|c}{ and value of $\omega$)}\\
\cline{3-5} &&
$n=11$,& $n=5$,& $n=11$,\\
&&
$\omega=0$&$\omega=0$&$\omega=0.8$\\
\hline
$\nu$&$D_{U_4}$&0.804(17)&0.968(19)&1.009(15)\\
$\nu$&$D_{U_2}$&0.789(16)&0.935(17)&0.977(13)\\
 \hline
$\overline{\nu}$&$D_{U_4},D_{U_2}$&0.796(12)&0.951(13)&0.993(10)\\
\hline
$\gamma/\nu$ &$\overline{\chi_{\rm max}}$                                        &1.845(11)&1.825(13)&1.748(16)\\
$\beta/\nu$ &$\overline{{\langle |{\mathcal M}|\rangle}_{\rm max}}$              &0.535(5 )&0.560(4 )&0.586(5 )\\
\hline
 $2\beta/\nu+\gamma/\nu$&$\overline{\chi_{\rm max}}$,$\overline{{\langle |{\mathcal M}|\rangle}_{\rm max}}$ &2.916(15)&2.945(15)&2.916(19)\\
\hline
\end{tabular}
\end{center}
\caption{\label{tab4} Values of the critical exponents obtained
from FSS of different observables via {\em averaging a}. 3rd and
5th columns: fit to all $n=11$ data points, 4th column: fit to the
5 last data points. }
\end{table}

First, we used the fit to the power laws (\ref{111b}) for all data
points. Resulting exponents are given in the third column of table
\ref{tab4}. Having a possibility to estimate $\nu$ from the
scaling of two different magnetic cumulants, $D_{U_2}$ and
$D_{U_4}$, we give in the table also an average value of $\nu$,
$\overline{\nu}$, that results from these estimates. However, such
a straightforward estimate of the exponents made for the lattice
sizes $L= 6 \div 96 $ may be not accurate enough, as far as the
correction-to-scaling or crossover effects may be present
(for an analysis of crossover effects in a similar context, see
Ref.~\cite{Berche02}).
Indeed, one can see
a bending of the curves in Fig. \ref{fig5} for $L=6\div12$. Taking
that correction to scaling is especially pronounced for the small
lattice sizes we decided to make a power law fit for five largest
lattice sizes only. The results for the exponents are given in the
fourth column of table \ref{tab4}. Another way to deal with the
correction-to-scaling phenomena is to explicitly take into account
the correction-to-scaling term in the fit via formula
(\ref{111d}). Using theoretical estimate for the
correction-to-scaling exponent of the 3d Ising model with
long-range-correlated disorder at the value of correlation
parameter (see Eq. (\ref{1})) $a=2$, $\omega(a=2)=0.8$
\cite{Blavatska01} and fitting all data points by the formula
(\ref{111d}) we arrive at the values of the exponents given in the
last column of table \ref{tab4}. An alternative way may be to keep
the value of the exponent $\omega$ as a fit parameter and fit all
data points to Eq. (\ref{111d}) to ensure the same value of the
leading exponent as those, obtained from the fit of data for five
largest lattice sizes. As we have checked such a procedure leads
to the values of exponent $\omega$ in the reasonable agreement
with the theoretical estimate $\omega=0.8$ \cite{Blavatska01}.

Comparing data for the exponents obtained by a power law fit
(\ref{111b}) for five last data points (five largest lattice
sizes) with the data obtained by a fit of all data points to the
expression that takes into account the correction-to-scaling
exponent, Eq. (\ref{111d}), we arrive to self-consistent results.
Indeed, as one can see form the table \ref{tab4}, the average
value of the correlation length exponent lies in the interval
$\overline{\nu}=0.95\div 0.99$, the other exponents are in the
range $\gamma/\nu= 1.75 \div$ 1.83, $\beta/\nu= 0.54 \div 0.58$.
Another check of the accuracy of the results obtained is the value
of the combination of the exponents $2\beta/\nu + \gamma/\nu$
which is to be equal to three by a hyperscaling relation $2\beta +
\gamma = d\nu$. These value is given in the last row of the table.

\subsubsection{\em Averaging b}\label{IVb2}

\begin{figure}[!ht]
\epsfxsize=7cm \centerline{\epsffile{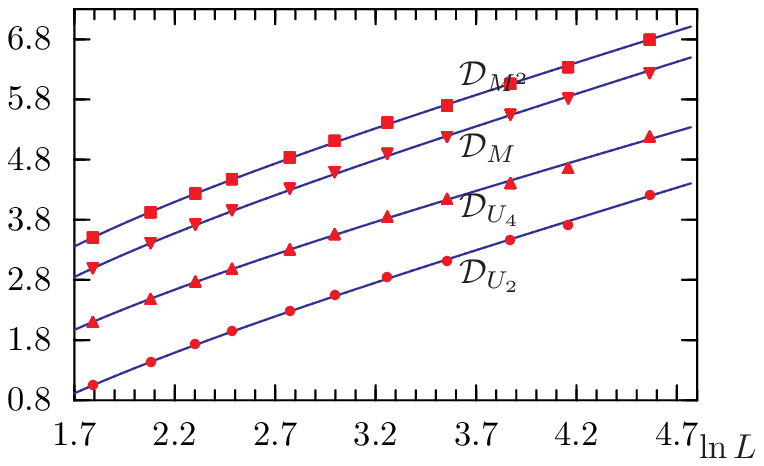}}
\caption{\label{fig55} Log-log plots for the maximum values of the
configurationally averaged derivatives of Binder cumulants
${\mathcal D}_{U_2}$ (discs), ${\mathcal D}_{U_4}$ (triangles up),
${\mathcal D}_{M}$ (triangles down) and ${\mathcal D}_{M^2}$
(squares). Solid line: data fit to the power law with the
correction-to-scaling, Eq. (\ref{111d}).}
\end{figure}

\begin{figure}[!ht]
\epsfxsize=6.5cm \epsffile{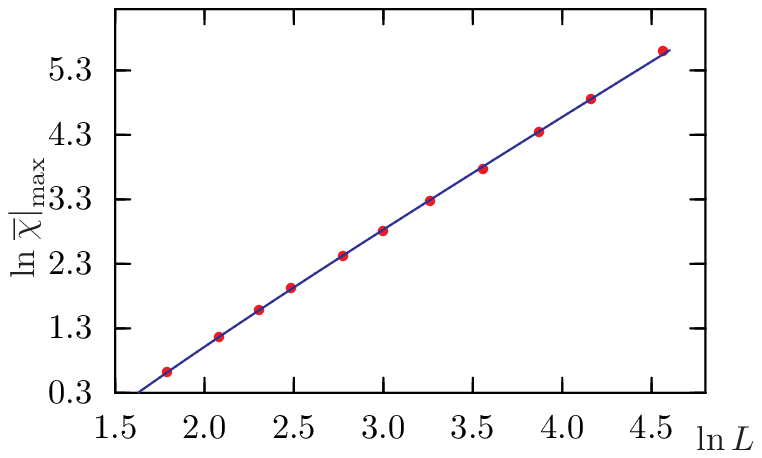} \epsfxsize=7cm
\epsffile{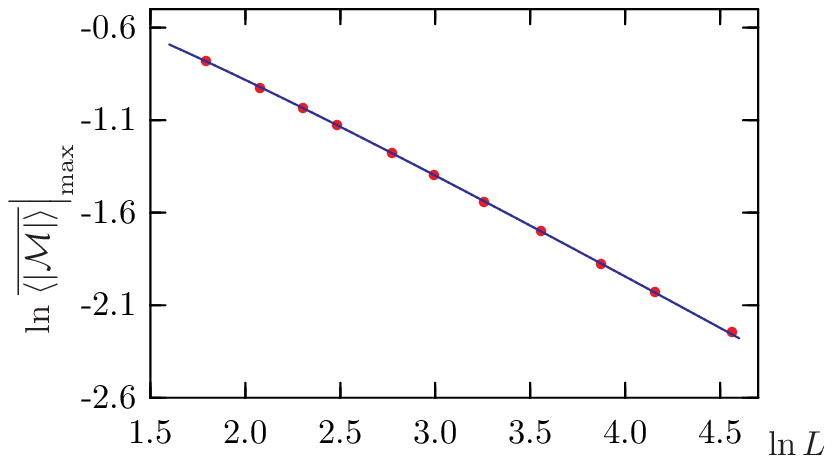} \centerline{(a)\hspace{8cm}(b) }
\caption{\label{fig75} Log-log plots for the configurationally
averaged maximum values of the magnetic susceptibility
$\left.\overline{\chi}\right|_{\rm max}$ and magnetisation
$\left.{\overline{\langle |{\mathcal M}|\rangle}}\right|_{\rm
max}$. Solid line: data fit to the power law with the
correction-to-scaling, Eq. (\ref{111d}).}
\end{figure}

\begin{figure}[!ht]
\epsfxsize=10cm \centerline{\epsffile{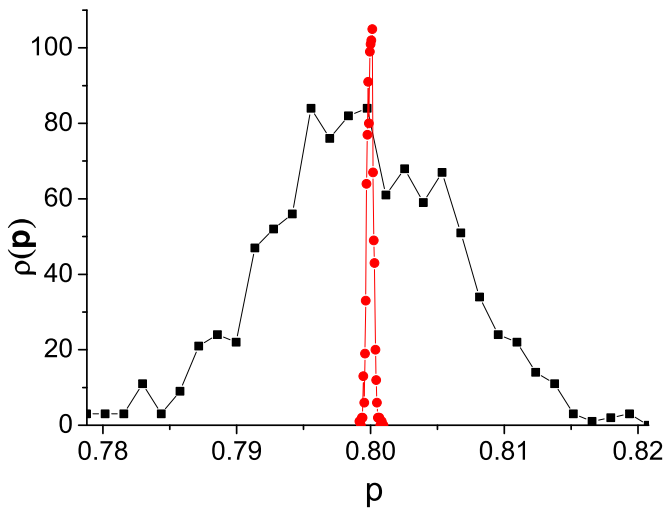}}
 \caption{Distribution function $\rho(p)$ of the magnetic spin concentration
$p$ for 'grand canonical disorder' and 'canonical disorder' in
comparison, ($L=96$).\label{hist}}
\end{figure}

Now let us carry out analysis being based on the same samples but to
perform an averaging we will use the procedure described above as an
{\em averaging b}. Results of the analysis are summarized in tables
\ref{tab5}, \ref{tab6} and Figures \ref{fig55}, \ref{fig75}.
Similarly as in the former subsection \ref{IVb1} we give in the
table \ref{tab5} the values of the observables that are used in the
FSS analysis. Now one can extract the correlation length critical
exponent $\nu$ from the FSS of maxima of four different quantities:
temperature derivatives of logarithm of magnetization and of its
square  ${\mathcal D}_{M}$,  ${\mathcal D}_{M^2}$ (\ref{111a}) and
of magnetic cumulants ${\mathcal D}_{U_4}$, ${\mathcal D}_{U_2}$
(\ref{111}). Corresponding plots are given in fig. \ref{fig55}.
Again as in the former subsection we used different ways to fit data
points to the power law dependence. On the one hand, we used simple
power law (\ref{111c}), on the other hand, the correction-to-scaling
was taken into account via formula (\ref{111d}).

\begin{sidewaystable}
\centering
\begin{tabular}{lllllll}
\hline\hline\\ & $\left.\overline{\chi}\right|_{\rm max}$& $\left.
{\overline{\langle |{\mathcal M}|\rangle}}\right|_{\rm max}$& ${\mathcal D}_{U_4}$& ${\mathcal D}_{U_2}$& ${\mathcal D}_{M}$& ${\mathcal D}_{M^2}$  \\
\hline
6   & $  1.850\pm0.003$& $ 0.4583  \pm0.0006 $ &  $ 8.178\pm0.001$ & $  2.850\pm0.001$  & $  19.958\pm0.007$  & $ 33.248 \pm0.013$  \\
8   & $  3.205\pm0.008$& $ 0.3974  \pm0.0006 $ &  $ 11.98\pm0.01$ & $  4.189\pm0.003$  & $  30.269\pm0.005$  & $ 50.630 \pm0.051$  \\
10  & $  4.872\pm0.016$& $ 0.3564  \pm0.0005 $ &  $ 15.99\pm0.03$ & $  5.641\pm0.012$  & $  41.245\pm0.059$  & $ 68.993 \pm0.159$  \\
12  & $  6.806\pm0.026$& $ 0.3245  \pm0.0002 $ &  $ 19.73\pm0.05$ & $  7.003\pm0.019$  & $  52.326\pm0.102$  & $ 87.714 \pm0.235$  \\
16  & $  11.27\pm0.054$& $ 0.2792  \pm0.0002 $ &  $ 27.29\pm0.11$ & $  9.807\pm0.042$  & $  74.989\pm0.238$  & $ 125.94 \pm0.47$  \\
20  & $  16.65\pm0.089$& $ 0.2474  \pm0.0002 $ &  $ 35.22\pm0.19$ & $  12.77\pm0.066$  & $  98.707\pm0.388$  & $ 165.89 \pm0.72$  \\
26  & $  26.32\pm0.157$& $ 0.2147  \pm0.0003 $ &  $ 47.06\pm0.24$ & $  17.20\pm0.10$  & $  133.97\pm0.63$  & $ 225.22 \pm1.14$  \\
35  & $  43.45\pm0.304$& $ 0.1837  \pm0.0004 $ &  $ 58.63\pm0.44$ & $  22.02\pm0.18$  & $  177.58\pm1.05$  & $ 299.58 \pm1.75$  \\
48  & $  77.51\pm0.583$& $ 0.1535  \pm0.0005 $ &  $ 84.83\pm0.08$ & $  31.81\pm0.22$  & $  256.23\pm1.55$  & $ 431.99 \pm2.67$  \\
64  & $  128.8\pm1.073$& $ 0.1314  \pm0.0006 $ &  $ 105.8\pm0.9$ & $  41.05\pm0.38$  & $  335.01\pm2.34$  & $ 565.50 \pm3.73$  \\
96  & $  269.0\pm2.457$& $ 0.1061  \pm0.0006 $ &  $ 178.4\pm1.2$ & $  67.75\pm0.49$  & $  588.48\pm3.01$  & $ 1012.7 \pm4.4$  \\
\hline
\end{tabular}
\caption{\label{tab5} MC simulation data obtained for 3d Ising
model with  long-range correlated disorder: {\em averaging b}.}
 \end{sidewaystable}

The values of the exponent $\nu$ for different fits are given in
the Table \ref{tab6} Having defined $\nu$ from the scaling of four
different quantities (${\mathcal D}_{U_4}$, ${\mathcal D}_{U_2}$,
${\mathcal D}_{M}$, ${\mathcal D}_{M^2}$) we find also an average
value on $\nu$, $\overline{\nu}$, and quote it in the table as
well. Similarly, from the log-log dependence of
$\left.\overline{\chi}\right|_{\rm max}$ and $\left.
{\overline{\langle |{\mathcal M}|\rangle}}\right|_{\rm max}$
(shown in fig. \ref{fig75}) we extract values of the exponents
$\gamma/\nu$ and $\beta/\nu$, see Eq. (\ref{111c}). Again the
values of the exponents that result from different fits are shown
in Table \ref{tab6}.

\begin{table}[htb]
\begin{center}
\begin{tabular}{l|l|l|l|l}
\hline\hline \multirow{4}{*}{Exponents}&
\multirow{4}{*}{Observables }& \multicolumn{3}{|c}{Data fit
(Number of
data points $n$}\\
& & \multicolumn{3}{|c}{ and value of $\omega$)}\\
\cline{3-5} &&
$n=11$,& $n=5$,& $n=11$,\\
&&
$\omega=0$&$\omega=0$&$\omega=0.8$\\
\hline
 $\nu$  &${\mathcal D}_{M^2}$ &0.796(20)&0.946(13)&0.989(11)\\
 $\nu$  &${\mathcal D}_{M}$   &0.754(24)&0.970(16)&0.977(14)\\
 $\nu$  &${\mathcal D}_{U_4}$ &0.827(20)&1.006(43)&1.051(18)\\
 $\nu$  &${\mathcal D}_{U_2}$ &0.816(18)&0.960(29)&1.002(15)\\
 \hline
$\overline{\nu}$&${\mathcal D}_{U_4},{\mathcal D}_{U_2},{\mathcal D}_{M},{\mathcal D}_{M^2}$&0.798(10)&0.971(14)&1.005(7)\\
\hline
$\gamma/\nu$ &$\left.\overline{\chi}\right|_{\rm max}$                                      &1.802(12)&1.779(14)&1.699(16)\\
$\beta/\nu$ &$\left.{\overline{\langle |{\mathcal M}|\rangle}}\right|_{\rm max}$            &0.525(4 )&0.540(6 )&0.567(5 )\\
\hline
 $2\beta/\nu+\gamma/\nu$&$\left.\overline{\chi}\right|_{\rm max}$,$\left.{\overline{\langle |{\mathcal M}|\rangle}}\right|_{\rm max}$  &2.852(14)&2.859(18)&2.833(19)\\
\hline
\end{tabular}
\end{center}
\caption{\label{tab6} Values of the critical exponents obtained
from FSS of different observables via {\em averaging b}. 3rd and
5th columns: fit to all $n=11$ data points, 4th column: fit to the
5 last data points.}
\end{table}

As it was  observed already  in section \ref{IVb1}, the strong
correction-to-scaling occurs for the small lattice sizes. Similar
phenomenon happens when one applies procedure of {\em averaging b}.
An evidence may serve the bending of the curves in Fig. \ref{fig55}
for the small $L$. Therefore we conclude that the most reliable
numerical data is obtained either from the FSS of five largest
lattices or from all data points but with an account of the
correction-to-scaling exponents. From table \ref{tab6} we conclude
that the exponent lays in the range $\overline{\nu}=0.97\div1.01$,
$\gamma/\nu=1.70\div1.78$, $\beta/\nu=0.54\div0.57$.

Tables \ref{tab4} and \ref{tab6} summarize results for the exponents
obtained via different averaging procedures ({\em a} and {\em b})
and via different fitting procedures for the same samples. Recall
that preparing the samples we have kept an average value of magnetic
sites concentration fixed and equal to $p=0.8$ with the dispersion
$\sigma^2=10^{-4}$. The next step in our analysis will be to check
how does the concentration fluctuations influence the values of the
(thermodynamical) critical exponents.

\subsection{Evaluation of the critical exponents. Canonical
disorder} \label{IVc}

Let us consider now the situation when the dispersion of
concentration $p$ is much smaller as it was taken in the former
subsection. that is, keeping the same average concentration of
magnetic sites $p=0.8$ let us consider much more narrow
distribution of its values for separate samples. For the study,
performed in this subsection we take the dispersion to be
$\sigma^2=10^{-7}$. Doing so we introduce 'canonical disorder' in
the spirit of Ref. \cite{Wiseman98}. As we will see below, this
step is also can serve as an separate independent check of the
values of the critical exponents we are interested in.

Similarly as it was described above, we perform an averaging over
disorder in two different ways, via {\em averaging a} and {\em
averaging b}. We extract the values of the exponents by the FSS
dependence of different observables as explicitly given in Eqs.
(\ref{111b}) - (\ref{111d}). Intermediate results and log-log plots
of these observables as functions of the lattice size are given in
Appendices A and B. The data for the exponents are summarized in the
tables \ref{tab8} and \ref{tab10} for averaging {\em a} and {\em b},
respectively.

\begin{table}[htb]
\begin{center}
\begin{tabular}{l|l|l|l|l}
\hline\hline \multirow{4}{*}{Exponents}&
\multirow{4}{*}{Observables }& \multicolumn{3}{|c}{Data fit
(Number of
data points $n$}\\
& & \multicolumn{3}{|c}{ and value of $\omega$)}\\
\cline{3-5} &&
$n=12$,& $n=5$,& $n=12$,\\
&&
$\omega=0$&$\omega=0$&$\omega=0.8$\\
\hline
$\nu$&$D_{U_4}$&0.841(15)&0.887(31)&0.989(15)\\
$\nu$&$D_{U_2}$&0.817(16)&0.904(19)&0.975(12)\\
 \hline
$\overline{\nu}$&$D_{U_4},D_{U_2}$&0.829(11)&0.896(18)&0.982(10)\\
\hline
$\gamma/\nu$ &$\overline{\chi_{\rm max}}$                                        &1.868(12)&1.999(33)&1.857(40)\\
$\beta/\nu$ &$\overline{{\langle |{\mathcal M}|\rangle}_{\rm max}}$              &0.526(2 )&0.527(2 )&0.538(9 )\\
\hline
 $2\beta/\nu+\gamma/\nu$&$\overline{\chi_{\rm max}}$, $\overline{{\langle |{\mathcal M}|\rangle}_{\rm max}}$ &2.920(13)&3.053(33)&2.933(44)\\
\hline
\end{tabular}
\end{center}
\caption{\label{tab8} Values of the critical exponents obtained
from FSS of different observables via {\em averaging a}. 3rd and
5th columns: fit to all $n=12$ data points, 4th column: fit to the
5 last data points.}
\end{table}

\begin{table}[htb]
\begin{center}
\begin{tabular}{l|l|l|l|l}
\hline\hline \multirow{4}{*}{Exponents}&
\multirow{4}{*}{Observables }& \multicolumn{3}{|c}{Data fit
(Number of
data points $n$}\\
& & \multicolumn{3}{|c}{ and value of $\omega$)}\\
\cline{3-5} &&
$n=12$,& $n=5$,& $n=12$,\\
&&
$\omega=0$&$\omega=0$&$\omega=0.8$\\
\hline
 $\nu$  &${\mathcal D}_{M^2}$ &0.807(13)&0.834(20)&0.932(12)\\
 $\nu$  &${\mathcal D}_{M}$   &0.764(18)&0.866(18)&0.956(12)\\
 $\nu$  &${\mathcal D}_{U_4}$ &0.855(7 )&0.958(13)&0.991(8 )\\
 $\nu$  &${\mathcal D}_{U_2}$ &0.841(18)&0.921(21)&1.031(17)\\
 \hline
$\overline{\nu}$&${\mathcal D}_{U_4},{\mathcal D}_{U_2},{\mathcal D}_{M},{\mathcal D}_{M^2}$& 0.817(7)&0.895(7)&0.978(6)\\
\hline
$\gamma/\nu$ &$\left.\overline{\chi}\right|_{\rm max}$                                       &1.800(12)&1.834(19)&1.731(27)\\
$\beta/\nu$ &$\left.{\overline{\langle |{\mathcal M}|\rangle}}\right|_{\rm max}$             &0.516(5 )&0.528(12)&0.562(4 )\\
\hline
 $2\beta/\nu+\gamma/\nu$&$\left.\overline{\chi}\right|_{\rm max}$,$\left.{\overline{\langle |{\mathcal M}|\rangle}}\right|_{\rm max}$ &2.832(16)&2.890(31)&2.855(28)\\
\hline
\end{tabular}
\end{center}
\caption{\label{tab10} Values of the critical exponents obtained
from FSS of different observables via {\em averaging b}. 3rd and
5th columns: fit to all $n=12$ data points, 4th column: fit to the
5 last data points.}
\end{table}

Comparing data of tables \ref{tab4} and \ref{tab6} with the data of
tables \ref{tab8} and \ref{tab10} ones can see that the values of
the exponents depend on the way the averaging over disorder was
performed ({\em averaging a} and {\em b}) as well as on the way the
disorder was prepared (grand canonical and canonical disorder). In
particular, passing from the grand canonical disorder to the
canonical disorder one observes an increase of $\gamma/\nu$ and
decrease of $\beta/\nu$. This difference in the exponents is more
pronounced for the {\em averaging a} (of order of 8 \%) and less
pronounced for the {\em averaging b} (of order of 2-4 \%). Deviation
in the exponent $\nu$ is of order of 5 \%. We further compare these
values with the other data available and make some conclusions in
the next section.

\section{Conclusions and outlook} \label{V}

As we explained in the introduction, our paper was inspired by
existing contradictory results about the critical behaviour of the
3d Ising model with long-range-correlated impurities. Whereas both
theoretical  and MC  studies
\cite{Weinrib83,eta,Prudnikov,Prudnikov05,Ballesteros99} agree about
the new universality class that arises in such a model, there exist
an essential disagreement between the estimates for the critical
exponents. For the value of the impurity correlation parameter $a=2$
they are summarized in table \ref{tab1}. The numerical data obtained
so far split into two groups giving essentially different results
for the exponents. To give an example, the predicted value for the
correlation length critical exponent $\nu$ deviates between these
groups within order of 30 \% ranging from $\nu=1$
\cite{Weinrib83,Ballesteros99} to $\nu\simeq0.71$
\cite{Prudnikov,Prudnikov05}.

To resolve existing contradictions, we performed MC simulations of
the 3d Ising model with long-range-correlated disorder in a form of
randomly oriented lines of non-magnetic sites (impurity lines).
Concentration of the impurities was taken to be $1-p=0.2$. We used
different ways to perform an averaging over disorder realizations,
referred as {\em averaging a} and {\em averaging b}. Moreover, we
used two different ways to prepare the samples: in one case we
allowed for a wide distribution of concentration (grand canonical
disorder) an the other case we made this distribution very narrow
(canonical disorder). Swendsen-Wang MC algorithm and a
finite-size-scaling analysis were applied to extract values of the
critical exponents that govern magnetic phase transition in such a
system. Our estimates for the critical exponents $\nu$,  $\beta$,
and $\gamma$ are given in tables \ref{tab4}, \ref{tab6}, \ref{tab8},
\ref{tab10}. As we have discussed above, the most reliable data fits
have been obtained by fitting the whole data for a given observable
set taking into account the correction-to-scaling exponent or by
fitting the data for the largest lattice sizes to the asymptotic
scaling behaviour. These results are summarized in table
\ref{tab11}.

\begin{sidewaystable}
\centering
\begin{tabular}{l|l|l|l|l|l|l|l|l|l}
\hline\hline \multirow{4}{*}{Exponents}& \multicolumn{4}{|c|}{grand
canonical disorder}& \multicolumn{4}{|c|}{canonical disorder}& \multirow{2}{*}{Averaged}\\
\cline{2-9} &\multicolumn{2}{|c|}{{\em averaging a}}&\multicolumn{2}{|c|}{{\em averaging b}}&\multicolumn{2}{|c|}{{\em average a}}&\multicolumn{2}{|c|}{{\em average b}}&\\
\cline{2-9} &$n=5$,&$n=11$,&$n=5$,&$n=11$,&$n=5$,&$n=12$,&$n=5$,&$n=12$,&\multirow{2}{*}{value}\\
&$\omega=0$&$\omega=0.8$&$\omega=0$&$\omega=0.8$&$\omega=0$&$\omega=0.8$&$\omega=0$&$\omega=0.8$&\\
\hline
$\overline{\nu}$        &0.951(13)&0.993(10)&0.971(14)&1.005(7 )&0.892(18)&0.982(10)&0.895(7 )&0.978(6 )&$0.958(4)$\\
$\gamma/\nu$            &1.825(13)&1.748(16)&1.779(14)&1.699(16)&1.999(33)&1.857(40)&1.834(19)&1.731(27)&$1.809(9)$\\
$\beta/\nu$             &0.560(4 )&0.586(5 )&0.540(6 )&0.567(5 )&0.527(2 )&0.538(9 )&0.528(12)&0.562(4 )&$0.551(2)$\\\hline
 $2\beta/\nu+\gamma/\nu$&2.945(15)&2.916(19)&2.859(18)&2.833(19)&3.053(33)&2.933(44)&2.890(31)&2.855(28)&$2.911(10)$\\
\hline
\end{tabular}
\caption{\label{tab11} Values of the critical exponents obtained via
{\em averaging a} and {\em averaging b} for the grand canonical and
for the canonical disorder. The last column gives averaged values.}
\end{sidewaystable}

One can see from the table, that different ways to perform (and to
analyze) the simulation lead to slightly different values of the
exponents. Whereas technical reasons for such deviations are
obvious, there is no physical reason for them. That is all cases
considered should correspond to the same asymptotic behaviour of
physical quantities. Moreover, if in a simulation only one case were
considered, there would be no way to see such deviation. Therefore,
we find it reasonable to look for the averaged values of the
exponents, as given in the last column of the table \ref{tab11}
\begin{equation}
\nu=0.958\pm 0.004,\ \gamma/\nu=1.809\pm 0.009,\ \beta/\nu=0.551\pm 0.002.
\end{equation}

Comparing our data quoted in table \ref{tab11} with those previously
obtained, table \ref{tab1}, one sees that our analysis leads to the
results differing from the existing so far and therefore on the
first sight does not make  the situation more clear. However, we
want to emphasize several items, opposing our results to previous
theoretical and numerical estimates.

As first discussed in Ref. \cite{Prudnikov}, a difference between
existing theoretical estimates for the exponents is caused by the
fact that the renormalization group analysis of Weinrib and Halperin
\cite{Weinrib83} was performed in the first order of the
perturbation theory with subsequent conjecture that the first order
result  $\nu=2/a$ is exact. Upcoming two-loop calculations at $d=3$
\cite{Prudnikov} gave numerical results for the exponents which
disagree with this conjecture and question it. Although one may
consider this result as a signal about non-trivial dependence of the
exponents on the disorder correlation parameter $a$ (as well as on
the order parameter dimension), one certainly should not take the
numbers obtained as a final estimates of the exponents. Indeed in
the renormalization group theory of disordered systems, and, more
general, of systems described by effective Hamiltonians of
complicated symmetry, one finds many examples when the two-loop
numerical estimates are essentially improved by higher-order
contributions (see e.g. reviews \cite{reviews} and references
therein). In support to this suggestion let us point to the fact,
that the two-loop theoretical estimates of Ref. \cite{Prudnikov}
bring about the negative sign for the pair correlation function
critical exponent $\eta$ (see table \ref{tab1}), whereas for the
$\phi^4$ theory the positiveness of $\eta$ follows from the
K\"allen-Lhemann decomposition. Therefore, our results for the
exponents support general scenario found in Ref. \cite{Prudnikov},
whilst the numerical discrepancy may be explained by possible
changes in the theoretical estimates due to the high-order
contributions.

Another question concerns discrepancies between the results of
numerical simulations of the 3d Ising model with long-range
correlated disorder at $a=2$ \cite{Ballesteros99,Prudnikov05}. In
Ref. \cite{Prudnikov05} the discrepancies were explained by the
differences in the objects of simulation: in the simulations of
Ballesteros et al. \cite{Ballesteros99} the impurity lines were
allowed to intersect opposite to the mutually avoiding impurity
lines considered in the simulations of Prudnikov et al.
\cite{Prudnikov}. One way to check whether the above difference in
the impurity distributions causes any influence on the magnetic
subsystem critical behaviour  is to analyze an asymptotics of the
function $g(r)$ (\ref{1}). As it follows from our analysis
\cite{Ivaneyko06b}, both distributions lead to the power law
asymptotics (\ref{1}) with close values of $a\simeq 2$. This
observation gives  an evidence of the fact that both disorder
distributions lead to the same values of the exponents. At least, if
the difference exists it can not be seen for the system sizes
considered. Moreover, if it exists it can not suffice to explain the
observed discrepancy between simulations \cite{Ballesteros99} and
\cite{Prudnikov05}. Indeed, the former simulations were performed
also for the appropriately distributed impurity sites (so-called
Gaussian disorder, see section \ref{I}), when the very notion of the
impurity lines and their intersections looses its sense.

Let us point another phenomenon where the results obtained here
may find possible interpretation. For liquids, a porous medium is
often better fitted to the extended long-range-correlated structures
\cite{Chakrabarti05,aerogels}. To give an example, experiments on
$^4$He embedded in silica aerogels and xerogels report a change in
the universality class \cite{helium} which signals about a presence
of long-range correlations between defects. Simple fluids immersed
in the porous media of certain type (determined by the
density-density pair correlation function of form of Eq. (\ref{1}))
should manifest critical behaviour in the universality class of the
3d Ising model with long-range-correlated disorder. Reported so far
values of the critical exponents for the 3d Ising model with defects
in a form of a porous medium \cite{Vasquez,MacFarland96} were rather
interpreted in terms of the random-site uncorrelated disorder (note
however a numerical agreement between the results of Ref.
\cite{MacFarland96} and our estimates). We hope that our simulations
will attract more attention to the interpretation of experimental
and numerical studies of criticality of liquids in porous medium in
terms of long-range-correlated disorder considered here.

Let us return back to the values of the exponents quoted in the last
column of table \ref{tab11}. From this data one gets the following
values of the magnetic susceptibility and magnetization critical
exponents: $\gamma=1.733(11)$, $\beta=0.528(3)$. Comparing these
values with the results of former MC simulation of Ballesteros and
Parisi \cite{Ballesteros99} (c.f. Table \ref{tab1}) one can see that
our result for $\beta$ nicely coincide however the value for
$\gamma$ differs of the order of 13 \%, the difference in $\nu$
being less dramatic. Note however that our results are in much worse
agreement with the simulations of Prudnikov et al.
\cite{Prudnikov05}. We do not see any obvious explanation for the
discrepancies between MC numerical estimates of the critical
behaviour of the 3d random Ising model with long-range-correlated
disorder obtained so far and those obtained in our study.
Eventually, if we are allowed to make a ``not very serious''
statement, we must confess that we simply ignore the reasons of the
discrepancy between previously available MC simulations and those
presented in this work. All three studies seem equally rigorous, but
eventually our analysis provides results for the critical exponents
which stand between former determinations, and this may be seen as a
wise behaviour!

The authors acknowledge useful discussions with Christophe Chatelain
and Wolfhard Janke. We are especially thankful to the unknown
referee for illuminating comments (that made our simulations even
more time consuming). Yu. H. acknowledges support of the FWF project
P 19583-PHY.

\section*{Appendix A}
\begin{figure}[!ht]
\epsfxsize=7cm \centerline{\epsffile{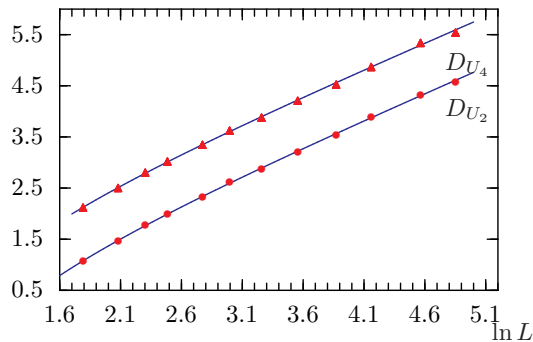}}
\caption{\label{fig56} Log-log plots for the maximum values of the
configurationally averaged derivatives of Binder cumulants $D_{U_2}$
(discs) and $D_{U_4}$ (triangles). Solid line: data fit to the power
law with the correction-to-scaling, Eq. (\ref{111d}).}
\end{figure}

\begin{figure}[!ht]
\epsfxsize=7cm \epsffile{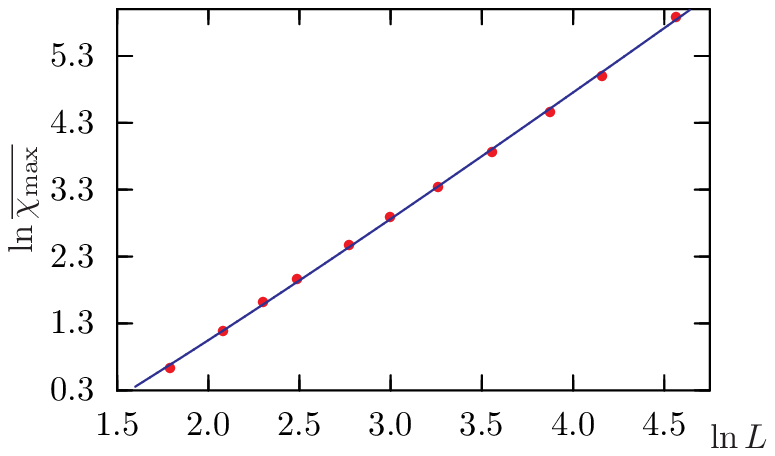} \epsfxsize=7cm \epsffile{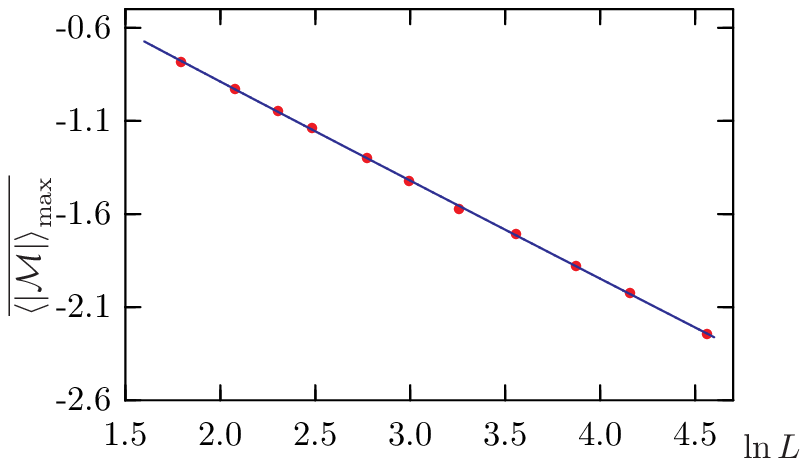}
\centerline{(a)\hspace{8cm}(b) } \caption{\label{fig76} Log-log
plots for the configurationally averaged maximum values of the
magnetic susceptibility $\overline{\chi_{\rm max}}$ and
magnetisation $\overline{{\langle|{\mathcal M}|\rangle}_{\rm max}}$.
Solid line: data fit to the power law with the
correction-to-scaling, Eq. (\ref{111d}).}
\end{figure}

In this appendix, we give data  obtained via {\em averaging a}
during MC simulations for the canonical disorder, as explained at
the beginning of section \ref{IVc}.  In table \ref{tab7} we give
numerical values of maxima of averaged quantities
$\overline{\chi_{\rm max}}$,  $\overline{{\langle |{\mathcal
M}|\rangle}_{\rm max}}$,  $D_{U_4}$, and $D_{U_2}$, Eq. (\ref{111}).
Log-log plots of the dependence of these quantities on the lattice
size $L$ are given in Figs. \ref{fig56} and \ref{fig76}.

\begin{sidewaystable}
\centering
\begin{tabular}{lllll}
\hline\hline
& $\overline{\chi_{\rm max}}$& $\overline{{\langle |{\mathcal M}|\rangle}_{\rm max}}$& $D_{U_4}$&$D_{U_2}$                                     \\
\hline
6   &   $ 1.875\pm0.003$& $0.4572\pm0.0002$&$8.323\pm0.007$ &$2.937\pm0.003$\\
8   &   $ 3.281\pm0.007$& $0.3947\pm0.0003$&$12.20\pm0.02$  &$4.338\pm0.007$\\
10  &   $ 5.067\pm0.014$& $0.3520\pm0.0003$&$16.47\pm0.04$  &$5.895\pm0.015$\\
12  &   $ 7.103\pm0.023$& $0.3197\pm0.0003$&$20.48\pm0.07$  &$7.354\pm0.024$\\
16  &   $ 11.89\pm0.05$ & $0.2726\pm0.0003$&$28.39\pm0.13$  &$10.28\pm0.05$\\
20  &   $ 17.94\pm0.08$ & $0.2415\pm0.0004$&$37.45\pm0.23$  &$13.59\pm0.09$\\
26  &   $ 28.09\pm0.14$ & $0.2072\pm0.0005$&$48.30\pm0.40$  &$17.80\pm0.15$\\
35  &   $ 47.45\pm0.29$ & $0.1810\pm0.0010$&$67.18\pm0.73$  &$24.50\pm0.26$\\
48  &   $ 86.64\pm0.58$ & $0.1531\pm0.0005$&$92.19\pm0.45$  &$34.56\pm0.24$\\
64  &   $ 147.8\pm1.1$  & $0.1319\pm0.0005$&$130.0\pm1.9$   &$48.67\pm0.77$\\
96  &   $ 360.0\pm2.5$  & $0.1060\pm0.0006$&$207.6\pm1.5$   &$75.81\pm0.61$\\
128 &   $ 624.2\pm4.0$  & $0.0915\pm0.0004$&$255.0\pm4.9$   &$96.69\pm2.14$\\
\hline
\end{tabular}
\caption{\label{tab7} MC simulation data obtained for 3d Ising model
with long-range correlated disorder: {\em averaging a}.}
\end{sidewaystable}

\section*{Appendix B}
\begin{figure}[!ht]
\epsfxsize=7cm \centerline{\epsffile{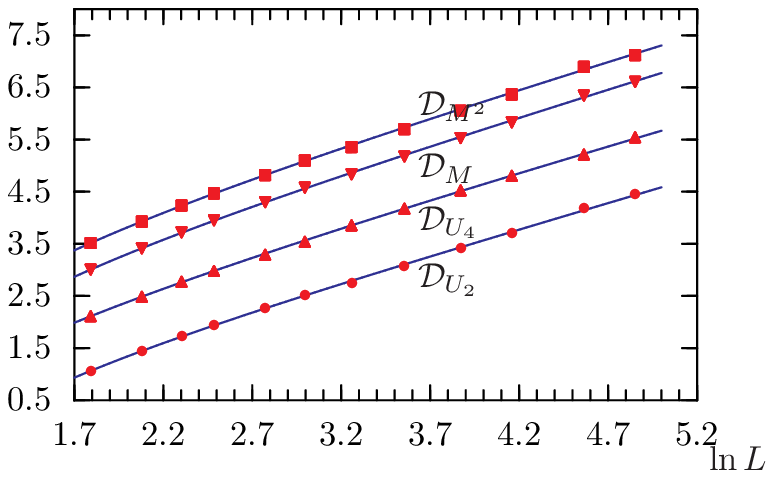}}
\caption{\label{fig57} Log-log plots for the maximum values of the
configurationally averaged derivatives of Binder cumulants
${\mathcal D}_{U_2}$ (discs), ${\mathcal D}_{U_4}$ (triangles up),
${\mathcal D}_{M}$ (triangles down) and ${\mathcal D}_{M^2}$
(squares). Solid line: data fit to the power law with the
correction-to-scaling, Eq. (\ref{111d}).}
\end{figure}

\begin{figure}[!ht]
\epsfxsize=6.5cm \epsffile{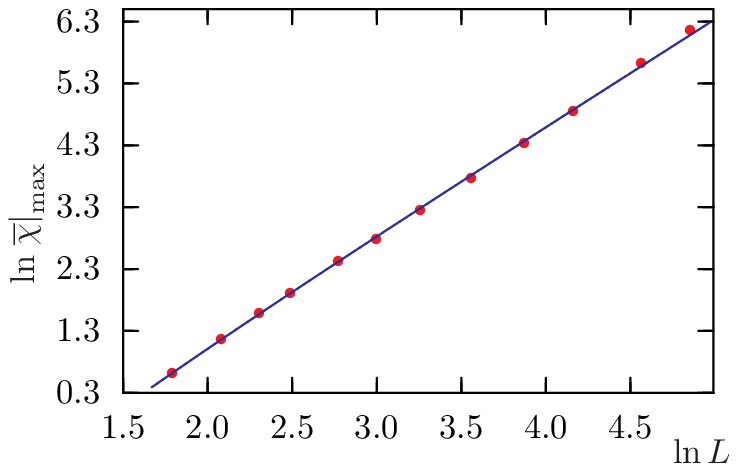} \epsfxsize=7cm
\epsffile{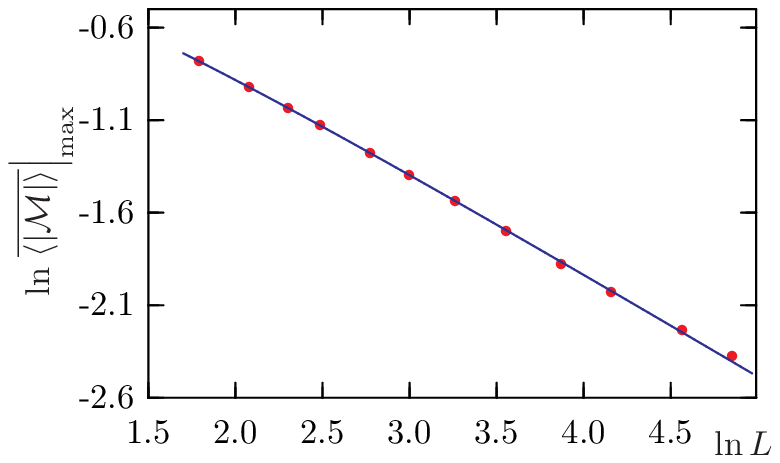} \centerline{(a)\hspace{8cm}(b) }
\caption{\label{fig77} Log-log plots for the configurationally
averaged maximum values of the magnetic susceptibility
$\left.\overline{\chi}\right|_{\rm max}$ and magnetisation
$\left.{\overline{\langle |{\mathcal M}|\rangle}}\right|_{\rm max}$.
Solid line: data fit to the power law with the
correction-to-scaling, Eq. (\ref{111d}).}
\end{figure}

In this appendix, we give data  obtained via {\em averaging b}
during MC simulations for the canonical disorder, as explained at
the beginning of section \ref{IVc}.  In table \ref{tab9} we give
numerical values of maxima of averaged quantities
$\left.\overline{\chi}\right|_{\rm max}$,  $\left.
{\overline{\langle |{\mathcal M}|\rangle}}\right|_{\rm max}$,
${\mathcal D}_{U_4}$,  ${\mathcal D}_{U_2}$,  ${\mathcal D}_{M}$,
${\mathcal D}_{M^2}$, Eqs. (\ref{111e}), (\ref{111a}).
 Log-log plots of the dependence of these
quantities on the lattice size $L$ are given in Figs. \ref{fig57}
and \ref{fig77}.

\begin{sidewaystable}
\centering
\begin{tabular}{lllllll}
\hline\hline & $\left.\overline{\chi}\right|_{\rm max}$& $\left.
{\overline{\langle |{\mathcal M}|\rangle}}\right|_{\rm max}$& ${\mathcal D}_{U_4}$& ${\mathcal D}_{U_2}$& ${\mathcal D}_{M}$& ${\mathcal D}_{M^2}$  \\
\hline
6   &   $ 1.848\pm0.003$ &$  0.4575\pm0.0003$ &$ 8.258\pm0.002$ &$ 2.896\pm0.001$ &$ 20.211\pm0.002$ &$ 33.623\pm0.018$ \\
8   &   $ 3.190\pm0.008$ &$  0.3959\pm0.0004$ &$ 11.99\pm0.01 $ &$ 4.217\pm0.005$ &$ 30.361\pm0.012$ &$ 50.713\pm0.060$ \\
10  &   $ 4.854\pm0.016$ &$  0.3545\pm0.0005$ &$ 16.00\pm0.03 $ &$ 5.647\pm0.012$ &$ 41.256\pm0.045$ &$ 69.040\pm0.134$ \\
12  &   $ 6.806\pm0.026$ &$  0.3241\pm0.0002$ &$ 19.67\pm0.06 $ &$ 6.979\pm0.022$ &$ 51.973\pm0.115$ &$ 87.095\pm0.263$ \\
16  &   $ 11.27\pm0.05 $ &$  0.2788\pm0.0002$ &$ 26.86\pm0.10 $ &$ 9.636\pm0.040$ &$ 73.721\pm0.229$ &$ 123.84\pm0.46 $ \\
20  &   $ 16.74\pm0.09 $ &$  0.2478\pm0.0004$ &$ 34.42\pm0.16 $ &$ 12.50\pm0.07 $ &$ 97.287\pm0.370$ &$ 163.54\pm0.68 $ \\
26  &   $ 25.71\pm0.16 $ &$  0.2150\pm0.0006$ &$ 47.22\pm0.15 $ &$ 15.69\pm0.10 $ &$ 125.62\pm0.62 $ &$ 211.95\pm1.05 $ \\
35  &   $ 43.62\pm0.33 $ &$  0.1869\pm0.0006$ &$ 65.00\pm0.05 $ &$ 21.65\pm0.15 $ &$ 176.67\pm0.75 $ &$ 299.21\pm1.61 $ \\
48  &   $ 76.92\pm0.58 $ &$  0.1276\pm0.0006$ &$ 94.12\pm1.35 $ &$ 30.61\pm0.44 $ &$ 249.97\pm2.93 $ &$ 420.41\pm2.72 $ \\
64  &   $ 129.2\pm1.0  $ &$  0.1319\pm0.0006$ &$ 122.0\pm0.4  $ &$ 40.69\pm0.36 $ &$ 340.26\pm2.68 $ &$ 581.36\pm3.83 $ \\
96  &   $ 279.0\pm2.6  $ &$  0.1074\pm0.0006$ &$ 183.6\pm1.5  $ &$ 65.51\pm0.61 $ &$ 568.72\pm2.92 $ &$ 986.67\pm4.78 $ \\
128 &   $ 470.9\pm4.7  $ &$  0.0939\pm0.0005$ &$ 253.0\pm4.9  $ &$ 86.69\pm2.14 $ &$ 740.37\pm8.36 $ &$ 1236.8\pm6.0  $ \\
\hline
\end{tabular}
\caption{MC simulation data obtained for 3d Ising model with
long-range correlated disorder: {\em averaging b}.\label{tab9}}
\end{sidewaystable} 

\end{document}